\documentclass{jpp}
\usepackage{graphicx}
\usepackage{hyperref}
\usepackage{amsmath}
\usepackage{csquotes}
\MakeOuterQuote{"}
\usepackage{bm}
\usepackage{overpic}
\usepackage{pgfplots}
\usepackage{tikz}
\usepackage[capitalise]{cleveref}  
\usepackage[acronym]{glossaries-extra}

\nonstopmode

\crefname{figure}{Fig.}{Figs.}
\crefname{equation}{Eq.}{Eqs.}
\crefname{section}{Sec.}{Secs.}

\glsdisablehyper
\setabbreviationstyle[acronym]{long-short}
\newacronym{ep}{EP}{enegetic particle}
\newacronym{mhd}{MHD}{magnetohydrodynamics}
\newacronym{pic}{PIC}{particle-in-cell}
\newacronym{iter}{ITER}{International Thermonuclear Experimental Reactor}
\newacronym{cfl}{CFL}{Courant–Friedrichs–Lewy}
\newacronym{rk4}{RK4}{4th order Runge-Kutta}
\newacronym{cpu}{CPU}{central processing unit}
\newacronym{gpu}{GPU}{graphics processing unit}
\newacronym{mpi}{MPI}{Message Passing Interface}
\newacronym{shm}{SHM}{MPI Shared Memory}
\newacronym{tae}{TAE}{toroidal Alfvén eigenmode}
\newacronym{rsae}{RSAE}{reversed shear Alfvén eigenmode}
\newacronym{gae}{GAE}{global Alfvén eigenmode}
\newacronym{cae}{CAE}{compressional Alfvén eigenmode}
\newacronym{flr}{FLR}{finite Larmor radius}
\newacronym{zlr}{ZLR}{zero Larmor radius}
\newacronym{fow}{FOW}{finite orbit width}
\newacronym{gs}{G-S}{Grad-Shafranov}
\newacronym{iaw}{IAW}{ion acoustic wave}
\newacronym{bae}{BAE}{beta-induced Alfvén eigenmode}
\newacronym{kaw}{KAW}{kinetic Alfvén wave}
\newacronym{lhd}{LHD}{Large Helical Device}

\shorttitle{Thermal ion kinetic effects and Landau damping in fishbone modes}
\shortauthor{Chang Liu et al.}

\title{Thermal ion kinetic effects and Landau damping in fishbone modes}

\author{Chang Liu\aff{1}
  \corresp{\email{cliu@pppl.gov}},
  Stephen C. Jardin\aff{1},
  Jian Bao\aff{2},
  Nikolai Gorelenkov\aff{1},
  Dylan P. Brennan\aff{1},
  James Yang\aff{1},
 \and Mario Podesta\aff{1}}

\affiliation{\aff{1}Princeton Plasma Physics Laboratory, Princeton, NJ 08540, USA
\aff{2}Institute of Physics, Chinese Academy of Sciences, Beijing 100190, China}

\begin{document}

\maketitle

\begin{abstract}
The kinetic-MHD hybrid simulation approach for macroscopic instabilities in plasmas can be extended to include the kinetic effects of both thermal ions and energetic ions. The new coupling scheme includes synchronization of density and parallel velocity between thermal ions and MHD, in addition to pressure coupling, to ensure the quasineutrality condition and avoid numerical errors. The new approach has been implemented in the kinetic-MHD code M3D-C1-K, and was used to study the thermal ion kinetic effects and Landau damping in fishbone modes in both DIII-D and NSTX. It is found that the thermal ion kinetic effects can cause an increase of the frequencies of the non-resonant $n=1$ fishbone modes driven by energetic particles for $q_\mathrm{min}>1$, and Landau damping can provide additional stabilization effects. A nonlinear simulation for $n=1$ fishbone mode in NSTX is also performed, and the perturbation on magnetic flux surfaces and the transport of energetic particles are calculated.
\end{abstract}

\section{Introduction}

The kinetic effects of thermal ions can become more important for physics studies targeting fusion reactors, given the large ion temperature ($T_i > 10$ keV) and the presence of high-energy alpha particles. In recent DIII-D experiments, it was observed that thermal ions can drive both high-frequency chirping modes\citep{du_multiscale_2021} and beta-induced Alfvén-acoustic eigenmodes (BAAEs)/low-frequency Alfvén modes\citep{gorelenkov_beta-induced_2009,choi_gyrokinetic_2021,ma_theoretical_2021}. These modes can lead to degradation of plasma confinement or even minor disruption. It is therefore necessary to incorporate these effects in numerical simulation models for ITER and future reactors.

However, the inclusion of thermal ion kinetic effects poses a challenge to classical hybrid simulation models\citep{todo_linear_1998,fu_global_2006,kim_impact_2008}, which combine \gls{pic} and the \gls{mhd} simulations. These models were developed for simulating the physics of \glspl{ep} or fast ions, which come from neutral beam injection. \glspl{ep} can have pressure or perpendicular current that is comparable to the thermal ions or electrons, but their density or momentum are often relatively small compared to the thermal ions. Their kinetic effects can then be included in the \gls{mhd} framework through pressure or current coupling, by including the $-\nabla\cdot\mathbf{P}$ or $\mathbf{J}\times\mathbf{B}$ term in the momentum equation, where $\mathbf{P}$ and $\mathbf{J}$ are calculated from \gls{ep} moments.

There are two major issues when extending this approach to thermal ions. First, since the \gls{mhd} density and momentum equation are derived by taking moments of ion and electron kinetic equations, when thermal ion kinetic equations are calculated separately, these \gls{mhd} equations become redundant and the error between the two approaches can lead to numerical issues or even parasitic modes. Second, as ions and electrons are calculated separately (one as kinetic particles and one as a fluid component), the parallel electric force between them must be calculated as a connection, which is often missing in ideal \gls{mhd} calculations or treated as a high-order two-fluid effect. 
 
In this paper, we describe a new kinetic-MHD coupling scheme, which is similar to the one used in the MEGA code for studying thermal ion kinetic effects in \gls{lhd} plasmas\citep{sato_effect_2019,sato_ion_2020}. In this scheme in addition to the coupling terms in the \gls{mhd} momentum equation, we have two more equations to connect the \gls{mhd} and kinetic parts. One is the synchronization of the parallel velocity between the kinetic ions and the \gls{mhd}, and the other is the synchronization of ion density. These two new equations are introduced to ensure quasi neutrality and avoid parasitic modes.

In this kinetic-MHD model, all the ions, including the thermal ones and the fast ones, are modeled as kinetic particles using the \gls{pic} method. The \gls{mhd} equation is in charge of calculating the evolution of fields and the electron pressure and temperature. The parallel electric field caused by separation of electrons and ions is also added in the ion kinetic equations. This scheme is implemented in the kinetic-MHD code M3D-C1-K \citep{liu_hybrid_2022}, which is based on the finite-element \gls{mhd} code M3D-C1\citep{ferraro_calculations_2009,jardin_multiple_2012}. It is found that the semi-implicit method introduced for solving \gls{mhd} equations\citep{jardin_multiple_2012} is helpful for stabilizing numerical instabilities after including ion kinetic terms, and the simulation can run with large timesteps to save computation time.

Using this new model, we investigate the kinetic effects of thermal ions, especially Landau damping, in the kinetic \gls{mhd} simulations. We first tested the new simulation model in an \gls{iaw} simulation, and achieved good agreement of mode frequency and damping rate with the theory. We then studied $n=1$ fishbone modes using DIII-D and NSTX equilibrium with $q_\mathrm{min}$ slightly larger than 1. This fishbone mode which is connected to the non-resonant (1,1) kink mode has been studied before using kinetic-MHD simulation without thermal ion kinetic effects\cite{brennan_energetic_2012,wang_linear_2013,shen_hybrid_2017,shen_hybrid_2020}.

We find that for simulation with only fast ions treated kinetically, the dominant $n=1$ mode has a transition from a classical fishbone mode to a \gls{bae} like mode as $q_\mathrm{min}$ increases, with a significant increase of mode frequency, which is consistent with the NIMROD simulation results\cite{brennan_energetic_2012}. After adding a kinetic treatment of the thermal ions, the frequencies of the fishbone modes increase and the growth rates decrease. The \gls{bae}-like mode branch becomes stable. These simulation results indicate that both of the two modes are strongly affected by the Landau damping effect from thermal ions which was not considered in previous kinetic-MHD simulations.
In addition, we did the nonlinear simulation for the fishbone mode to study the mode saturation and the effects on particle transport.  

The paper is organized as follows. In \cref{sec:kinetic-mhd-model} we discuss the kinetic-MHD coupling scheme including the thermal ions, with density and parallel velocity synchronization. In \cref{sec:iaw} we present the simulation results of \glspl{iaw} using the new scheme, and compare the Landau damping rates with theoretical results. In \cref{sec:d3d-fishbone} we show the numerical simulation of $n=1$ fishbone modes in a DIII-D equilibrium without and with the thermal ions kinetic effects.
In \cref{sec:nstx-fishbone} we did similar simulations in NSTX scenarios, using a more realistic \gls{ep} distribution. In \cref{sec:nstx-nonlinear} we show the results of a nonlinear simulation of fishbones in NSTX, focusing on the mode saturation behavior.
Finally, the summary is given in \cref{sec:summary}.

\section{Kinetic-MHD model with thermal ion kinetic effects}
\label{sec:kinetic-mhd-model}

In this section we present the kinetic-MHD model implemented in the M3D-C1-K code. The code was initially developed as a kinetic module for the extended \gls{mhd} code M3D-C1\citep{jardin_multiple_2012} using a pressure coupling scheme, which is similar to other kinetic-MHD codes like M3D-K\citep{fu_global_2006} and NIMROD\citep{kim_impact_2008}. The particle equation of motion, $\delta f$ equations, and coupling scheme are described in \citet{liu_hybrid_2022}. In the new version of M3D-C1-K, we treat both the thermal ions and fast ions as kinetic particles and calculate their dynamics using the \gls{pic} method, and electrons are treated using a fluid model. The \gls{mhd} equations with kinetic coupling are as follows,
\begin{align}
	\label{eq:dvperpdt}
	\rho\left[\frac{\partial \mathbf{v}_\perp}{\partial t}+\left(\mathbf{v}_\perp+\mathbf{v}_\parallel\mathbf{b}\right)\cdot\nabla\mathbf{v}_\perp\right]=&\mathbf{J}\times\mathbf{B}-\nabla_\perp p_e-\nabla_\perp\cdot\left[P_{i\parallel}\mathbf{b}\mathbf{b}+P_{i\perp}\left(\mathbf{I}-\mathbf{b}\mathbf{b}\right)\right]\nonumber\\
	&-\nabla_\perp\cdot\left[P_{f\parallel}\mathbf{b}\mathbf{b}+P_{f\perp}\left(\mathbf{I}-\mathbf{b}\mathbf{b}\right)\right]+\nu\nabla^2\mathbf{v}_\perp,
\end{align}
\begin{equation}
	\mathbf{J}=\frac{1}{\mu_0}\nabla\times\mathbf{B},
\end{equation}
\begin{equation}
	\frac{\partial \mathbf{B}}{\partial t}=-\nabla\times\mathbf{E},
\end{equation}
\begin{equation}
	\label{eq:ohmslaw}
	\mathbf{E}=-\mathbf{v}_\perp\times\mathbf{B}+\eta \mathbf{J}.
\end{equation}
Here $\rho$ is the total ion mass density, $\mathbf{v}_\parallel$ and $\mathbf{v}_\perp$ are the \gls{mhd} velocity parallel and perpendicular to the magnetic field $\mathbf{B}$, $\mathbf{J}$ is the plasma current density, $\mathbf{E}$ is the electric field, $\nu$ and $\eta$ are viscosity and resistivity coefficients. $\mathbf{P}_\parallel$ and $\mathbf{P}_\perp$ are the parallel and perpendicular components of the ion pressure tensor, and the subscript $i$ and $f$ represents thermal ion and fast ions. The electron pressure $p_e$ can either be calculated with the convection and diffusion terms,
\begin{align}
	\label{eq:pe}
	\frac{\partial p_e}{\partial t}+\left(\mathbf{v}_\perp+\mathbf{v}_\parallel\mathbf{b}\right)\cdot\nabla p_e=&-\gamma_e p_e\nabla\cdot\left[\left(\mathbf{v}_\perp+\mathbf{v}_\parallel\mathbf{b}\right)\right]+n_e\nabla\cdot\left[\kappa_\perp\mathbf{I}+\kappa_\parallel\mathbf{b}\mathbf{b}\right]\cdot\nabla\left(\frac{p_e}{n_e}\right),
\end{align}
or as a product of density and electron temperature ($p_e=n_eT_e$). The temperature can be calculated separately
\begin{align}
	\label{eq:te}
	\frac{\partial T_e}{\partial t}+\left(\mathbf{v}_\perp+\mathbf{v}_\parallel\mathbf{b}\right)\cdot\nabla T_e=&-\left(\gamma_e-1\right)T_e\nabla\cdot\left[\left(\mathbf{v}_\perp+\mathbf{v}_\parallel\mathbf{b}\right)\right]+\nabla\cdot\left[\kappa_\perp\mathbf{I}+\kappa_\parallel\mathbf{b}\mathbf{b}\right]\cdot\nabla T_e,
\end{align}
where $\gamma_e$ is the electron specific heat ratio, $\kappa_\parallel$ and $\kappa_\perp$ are the parallel and perpendicular heat transport coefficients.

The kinetic ion orbit follows the guiding-center equation of motion,
\begin{equation}
	\label{eq:rk4-1}
	\frac{d\mathbf{X}}{dt}=\frac{1}{B^\star}\left[V_\parallel \mathbf{B}^\star-\mathbf{b}\times \left(\mathbf{E}-\frac{\mu}{q}\nabla B\right)\right],
\end{equation}
\begin{equation}
	\label{eq:rk4-2}
	m \frac{d V_\parallel}{dt}=\frac{1}{B^\star}\mathbf{B}^\star\cdot \left(q\mathbf{E}-\mu\nabla B\right),
\end{equation}
where
\begin{equation}
	\mathbf{B}^\star=\mathbf{B}+\frac{m V_\parallel}{q}\nabla\times\mathbf{b},
\end{equation}
\begin{equation}
	\label{eq:rk4-4}
	B^\star=\mathbf{B}^\star\cdot \mathbf{b}.
\end{equation}
\begin{equation}
	\label{eq:estar}
	\mathbf{E}^\star=\mathbf{E}-\frac{1}{n_e e}\nabla_\parallel p_e.
\end{equation}
Here $\mathbf{X}$ is the particle guiding center location, $V_\parallel$ the the particle parallel velocity. $m$ is the ion mass. $\mu=mV_\perp^2/2B$ is the magnetic moment. $\mathbf{b}=\mathbf{B}/B$. Note that the electric field $\mathbf{E}^\star$ has an additional term in the parallel direction, $-\nabla_\parallel p_e/n_e e$. This term is obtained by ignoring the inertial term in the electron momentum equation in the parallel direction.

Assuming that the ion distribution is given as a function of location $\mathbf{X}$, energy $\mathcal{E}=(1/2)mV^2$ and pitch angle $\xi=V_\parallel/V$, the weight equation for $\delta f$ calculation is obtained from drift kinetic equations,
\begin{equation}
	\frac{d w}{dt}=-\alpha\left[\left(\frac{d\mathbf{X}}{dt}\right)_1\cdot\nabla+\left(\frac{d\mathcal{E}}{dt}\right)_1\frac{\partial}{\partial \mathcal{E}}+\left(\frac{d\xi}{dt}\right)_1\frac{\partial}{\partial \xi}\right]\ln f_0,
\end{equation}
where
\begin{equation}
	\left(\frac{d\mathbf{X}}{dt}\right)_1=\frac{\mathbf{E}\times\mathbf{B}}{B^2}+V_\parallel\delta\mathbf{b},
\end{equation}
\begin{equation}
	\left(\frac{d\mathcal{E}}{dt}\right)_1=\left[V_\parallel \mathbf{b}+\frac{mV_\parallel^2}{qB}\mathbf{b}\cdot\nabla\times\mathbf{b}+\frac{\mu}{qB}\mathbf{b}\times\nabla B\right]\cdot q\mathbf{E}^\star,
\end{equation}
\begin{equation}
	\left(\frac{d\mathcal{\xi}}{dt}\right)_1=\frac{1}{V}\frac{d V_\parallel}{dt}-\frac{2V_\parallel}{mV^3}\frac{d\mathcal{E}}{dt},
\end{equation}
\begin{equation}
	\left(\frac{d V_\parallel}{dt}\right)_1=\left[\mathbf{b}+\frac{mV_\parallel}{qB}\nabla\times\mathbf{b}\right]\cdot\frac{q}{m}\mathbf{E}^\star+\delta\mathbf{b}\cdot\left(-\frac{\mu}{m}\nabla B\right).
\end{equation}
$\alpha=1$ for linear calculation and $\alpha=1-w$ for nonlinear. Alternatively, one can calculate the difference between the results of the equations of motion with equilibrium and perturbed fields to obtain these $\left(\dots\right)_1$ terms.

Finally, the density, parallel velocity and pressure for thermal (subscript $i$) and fast (subscript $f$) ions used in the \gls{mhd} equations are calculated from particle deposition on fields,
\begin{equation}
	\delta n_{i,f}(\mathbf{x})=\sum_{k_{i,f}} \left(w_{k_{i,f}}+\frac{\delta B_{\parallel}}{B_0^\star}\right) S\left(\mathbf{x}-\mathbf{x}_{k_{i,f}}\right),
\end{equation}
\begin{equation}
	\delta \rho=m_i\delta n_i+m_f \delta n_f,
\end{equation}
\begin{equation}
	\delta n_e=Z_i\delta n_i+Z_f \delta n_f,
\end{equation}
\begin{align}
	\label{eq:vpar}
	\delta v_\parallel(\mathbf{x})=\frac{1}{n_{e0}+\delta n_e}&\left[\sum_{k_i} Z_i V_{\parallel,k_i} \left(w_{k_i}+\frac{\delta B_{\parallel}}{B_0^\star}\right) S\left(\mathbf{x}-\mathbf{x}_k\right)-Z_i n_{i0}v_{\parallel i,0}\right.\nonumber\\
	+&\left.\sum_{k_f} Z_f V_{\parallel,k_f} \left(w_{k_f}+\frac{\delta B_{\parallel}}{B_0^\star}\right) S\left(\mathbf{x}-\mathbf{x}_k\right)-Z_f n_{f0}v_{\parallel f,0}\right],
\end{align}
\begin{equation}
	\label{eq:ppar}
	\delta P_{\parallel i,f}(\mathbf{x})=\sum_{k_{i,f}} m_{i,f} V_{\parallel,k_{i,f}}^2 \left(w_{k_{i,f}}+\frac{\delta B_{\parallel}}{B_0^\star}\right) S\left(\mathbf{x}-\mathbf{x}_{k_{i,f}}\right),
\end{equation}
\begin{equation}
	\label{eq:pperp}
	\delta P_{\perp i,f}(\mathbf{x})=\sum_{k_{i,f}} \mu_{k_{i,f}} B_0 \left(w_{k_{i,f}}+\frac{\delta B_{\parallel}}{B_0^\star}+\frac{\delta B_{\parallel}}{B_0}\right)  S\left(\mathbf{x}-\mathbf{x}_{k_{i,f}}\right)
\end{equation}
Here $S$ is the particle shape function, $k_{i,f}$ are index of particles, $m_i$ and $m_f$ are the ion mass, $Z_i$ and $Z_f$ are ion effective charge,  $v_{\parallel 0}$ is the parallel velocity of ion equilibrium distribution $f_0$. The details of the implementation of the particle deposition calculation can be found in \citet{liu_hybrid_2022}. For simulation with \gls{flr} effects, both the field evaluation and the particle deposition need to take into account average along the gyro orbit.

The kinetic-MHD scheme here is similar to that implemented in MEGA\citep{sato_effect_2019,sato_ion_2020}, except that we use pressure coupling while MEGA uses current coupling. As pointed out in \citet{liu_hybrid_2022}, the two kinds of coupling schemes are equivalent for calculating $v_\perp$. Comparing with the pressure coupling scheme with thermal ions in \citet{park_plasma_1999}, we have additional equations for synchronization of ion density and parallel velocity with \gls{mhd} fields, whereas in \citet{park_plasma_1999} these quantities are calculated by the \gls{mhd} equations. For example, the parallel velocity is solved as
\begin{align}
	\label{eq:dvpardt}
	\rho\left[\frac{\partial v_\parallel}{\partial t}+\left(\mathbf{v}_\perp+v_\parallel\mathbf{b}\right)\cdot\nabla v_\parallel\right]=&-\nabla_\parallel\cdot\left[P_{i\parallel}\mathbf{b}\mathbf{b}+P_{i\perp}\left(\mathbf{I}-\mathbf{b}\mathbf{b}\right)\right]\nonumber\\
	&-\nabla_\parallel\cdot\left[P_{f\parallel}\mathbf{b}\mathbf{b}+P_{f\perp}\left(\mathbf{I}-\mathbf{b}\mathbf{b}\right)\right]-\nabla_\parallel p_e.
\end{align}
If assuming small ion energy and ignoring the gradient and curvature drifts, one can verify that \cref{eq:dvpardt} can be obtained by taking the moment of kinetic equations. Here $\nabla_\parallel p_e$ is derived from the parallel electric field term in \cref{eq:estar} which is used in particle weight equations. In principle, the pressure coupling scheme in \citet{park_plasma_1999} is equivalent to our coupling scheme and the simulation results should agree with each other, though the calculation of ion density and $v_\parallel$ in \gls{mhd} equations are redundant.

However, we find that in M3D-C1, due to the fact that the fluid equations and kinetic equations are solved subsequently in one timestep, in the simulation using the scheme of \citet{park_plasma_1999}, the difference between $\delta n$ and $\delta v_\parallel$ from the fluid and kinetic equations can increase with time, which violates the quasi neutrality condition and lead to parasitic modes that overwhelm the numerical result. We thus replace the two \gls{mhd} equations with the synchronization schemes instead to avoid these issues. The perpendicular momentum equation (\cref{eq:dvperpdt}) of \gls{mhd} is kept as it provides additional information for kinetic simulation. In fact, \cref{eq:dvperpdt} can be regarded as a decomposition of plasma perpendicular current, in which $\rho \partial \mathbf{v}_\perp/\partial t$ term represents the polarization current, and the pressure terms represent the drift and magnetization current. 

The ion's parallel velocity is used in the electron pressure (\cref{eq:pe}) and temperature equation (\cref{eq:te}), assuming that the electrons and ions are moving together. However, when considering two-fluid effects, there is a difference between the electron and ion velocities due to the parallel current. This difference can leads to correction terms in \cref{eq:pe} and \cref{eq:te} that are proportional to $d_i/L$ ($d_i$ is the ion skin depth), and thus can be considered as a two-fluid term. These two-fluid terms are ignored in the current simulation model as we are dealing with long-wavelength modes. The additional parallel electric field used in \cref{eq:estar} is not included in the Ohm’s law (\cref{eq:ohmslaw}) for the same reason. These two-fluid effects can be important for high-$k$ mode simulations like \gls{kaw}, and will be studied in future.

\section{Numerical simulation of ion acoustic wave}
\label{sec:iaw}

In this section we test the new version of M3D-C1-K in an \gls{iaw} simulation. The oscillation of \glspl{iaw} can be easily simulated using a \gls{mhd} code. However, for $T_i \sim T_e$, IAW will be strongly damped due to parallel Landau damping, which can only be simulated by including the kinetic effects of thermal ions.

In the simulation we treat thermal ions as kinetic particles and electrons as a fluid component. Since \gls{iaw} is an electrostatic mode, we only keep the \gls{mhd} equation \cref{eq:pe} and ignore the $v_\perp$ terms. The \gls{mhd} equation and the particle equation of motion are limited to one-dimensional along the wave vector. For electron pressure we choose the electron heat capacity ratio $\gamma_e=1$ assuming they are isothermal. The parallel velocity is initialized like a sinusoidal function with a fixed wave number $k$, which can drive perturbations on $p_e$ and $p_i$ and lead to a standing wave. The electron and ion density are the same assuming $Z_{\mathrm{eff}}=1$, and the ion temperature is set to be a fraction of electron temperature.

\begin{figure}
	\centerline{\includegraphics[width=0.45\linewidth]{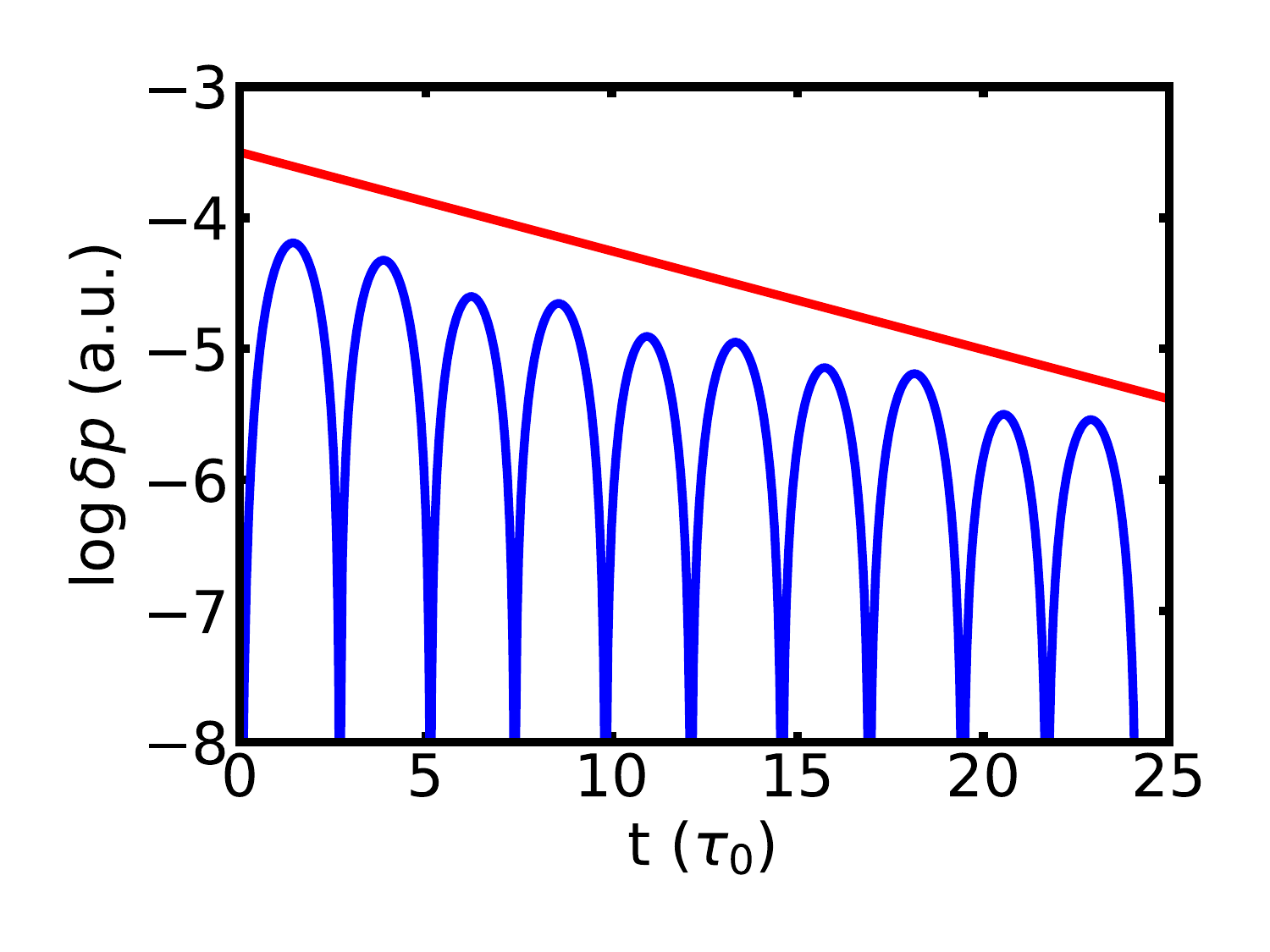}}
	\caption{Blue line is the time signal of $\delta p_e$ from \gls{iaw} simulation with $T_i=0.2 T_e$. Red line shows a mode damping trend with a rate calculated from \cref{eq:damping-landau}. The time unit $\tau_0=1/\omega_0$.}
	\label{fig:landau-damping-signal}
\end{figure}

For \gls{mhd}-only simulation, we find that \gls{iaw} gives oscillations of $\delta p$ and $\delta v_\parallel$ with little damping. With ion kinetic effects, the oscillation experiences damping as shown in \cref{fig:landau-damping-signal}. We find that the mode damping rate is consistent with the theoretical Landau damping rate of \gls{iaw} which is shown as the red line. Note that for $T_i/T_e>0.1$, the Landau damping rates $\gamma_{LD}$ of \glspl{iaw} become comparable to $\omega$ and the perturbative calculation is not accurate. In these cases $\gamma_{LD}$ should be solved numerically from the plasma dispersion function $Z(\zeta)$. Here we use empirical functions to calculate the frequencies and the Landau damping rates of \glspl{iaw} for $0.1<T_i/T_e<1$ from \citet{chen_introduction_2013},

\begin{equation}
	\label{eq:omega-landau}
	\omega=\omega_0\sqrt{1+\frac{3T_i}{T_e}}
\end{equation}
\begin{equation}
	\label{eq:damping-landau}
	\frac{\gamma_{LD}}{\omega}=1.1*\left(\frac{T_i}{T_e}\right)^{7/4}\exp\left[-\left(\frac{T_i}{T_e}\right)^2\right]
\end{equation}
where $\omega_0=k\sqrt{k_B T_e/m_i}$ and $k_B$ is the Boltzmann constant. \cref{eq:omega-landau} is calculated by assuming the ion heat capacity ratio $\gamma_i=3$ since they only suffer one-dimension compression\citep{mckinstrie_accurate_1999,chen_introduction_2013}.

\begin{figure}
	\begin{center}
		\begin{overpic}[width=0.4\textwidth]{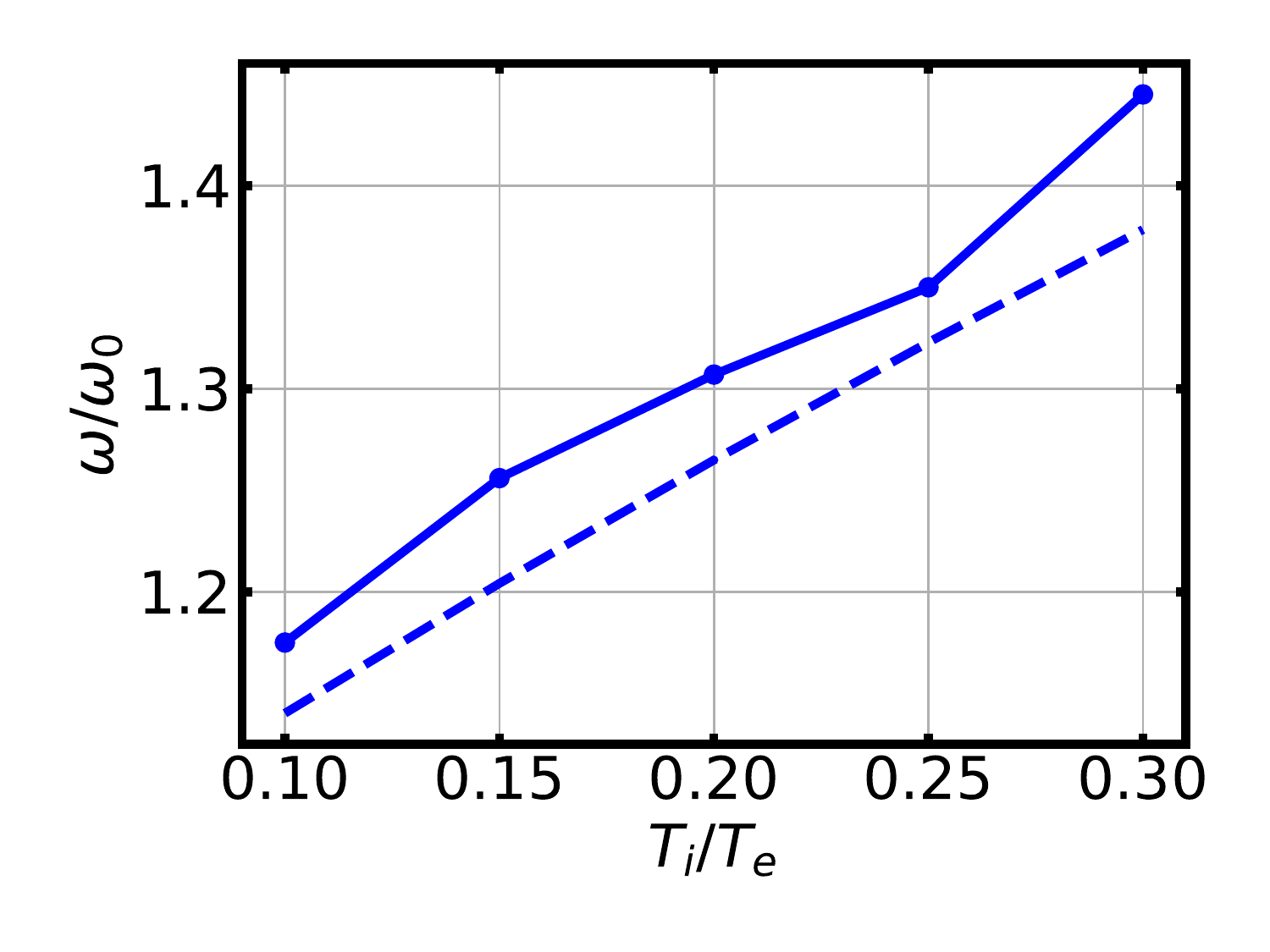}
			\put(83,20) {\textsf{(a)}}
		\end{overpic}
		\begin{overpic}[width=0.4\textwidth]{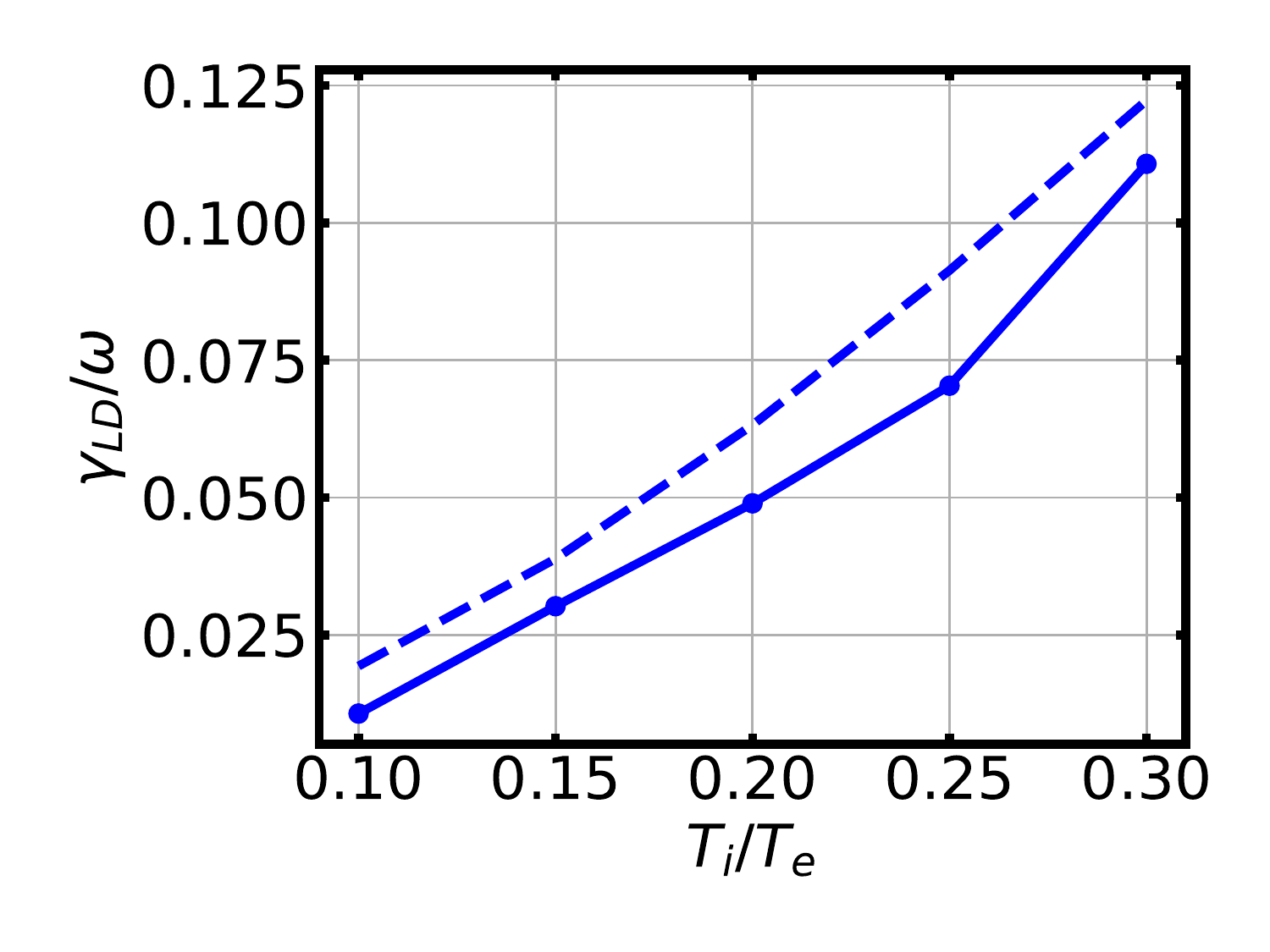}
			\put(83,20) {\textsf{(b)}}
		\end{overpic}
	\end{center}
	\caption{Frequencies (a) and damping rates (b) from \gls{iaw} simulation for different values of $T_i/T_e$. The dashed lines are the theoretical results calculated from \cref{eq:omega-landau} and \cref{eq:damping-landau}.}
	\label{fig:landau-damping-rate}
\end{figure}

\cref{fig:landau-damping-rate} shows the results of \gls{iaw} frequencies and damping rates from the M3D-C1-K simulation for different values of $T_i/T_e$. For small $T_i/T_e$, the damping rate is zero and the frequency is close to the \gls{mhd}-only result $\omega_0$. For $T_i/T_e>0.3$, the damping rate becomes comparable to the frequency indicating \glspl{iaw} are strongly damped. Both $\omega$ and $\gamma_{LD}$ are close to the theoretical results, indicating that the new kinetic-MHD model successfully captures the Landau damping physics.

\begin{figure}
	\centerline{\includegraphics[width=0.45\linewidth]{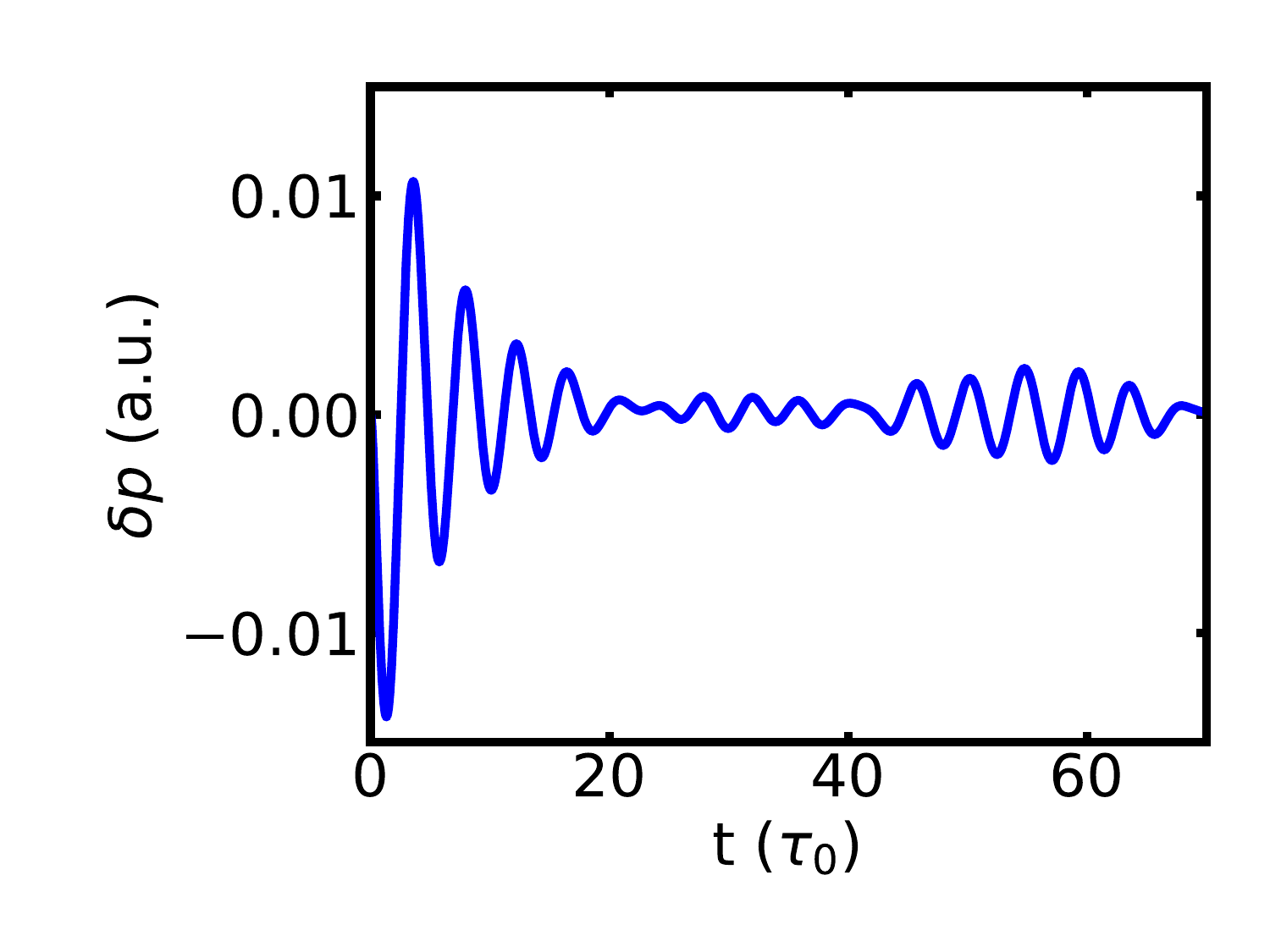}}
	\caption{Time evolution of $\delta p_e$ from IAW simulation with $T_i=0.3 T_e$, showing echos of oscillation after the mode damped.}
	\label{fig:landau-damping-echo}
\end{figure}

We find that for large $T_i/T_e$, the mode can have echos after being significantly damped, as shown in \cref{fig:landau-damping-echo}. This phenomenon is a typical nonlinear behavior for the Landau damped mode, which is different from dissipative damping\citep{kadomtsev_landau_1968}. This effect indicates that although the mode is damped due to phase mixing, the particles still retain some "memory" of the preceding oscillation which can be reflected as echos at later times.

\section{Linear simulation of fishbone modes in DIII-D}
\label{sec:d3d-fishbone}

\begin{figure}
	\begin{center}
		\begin{overpic}[width=0.4\textwidth]{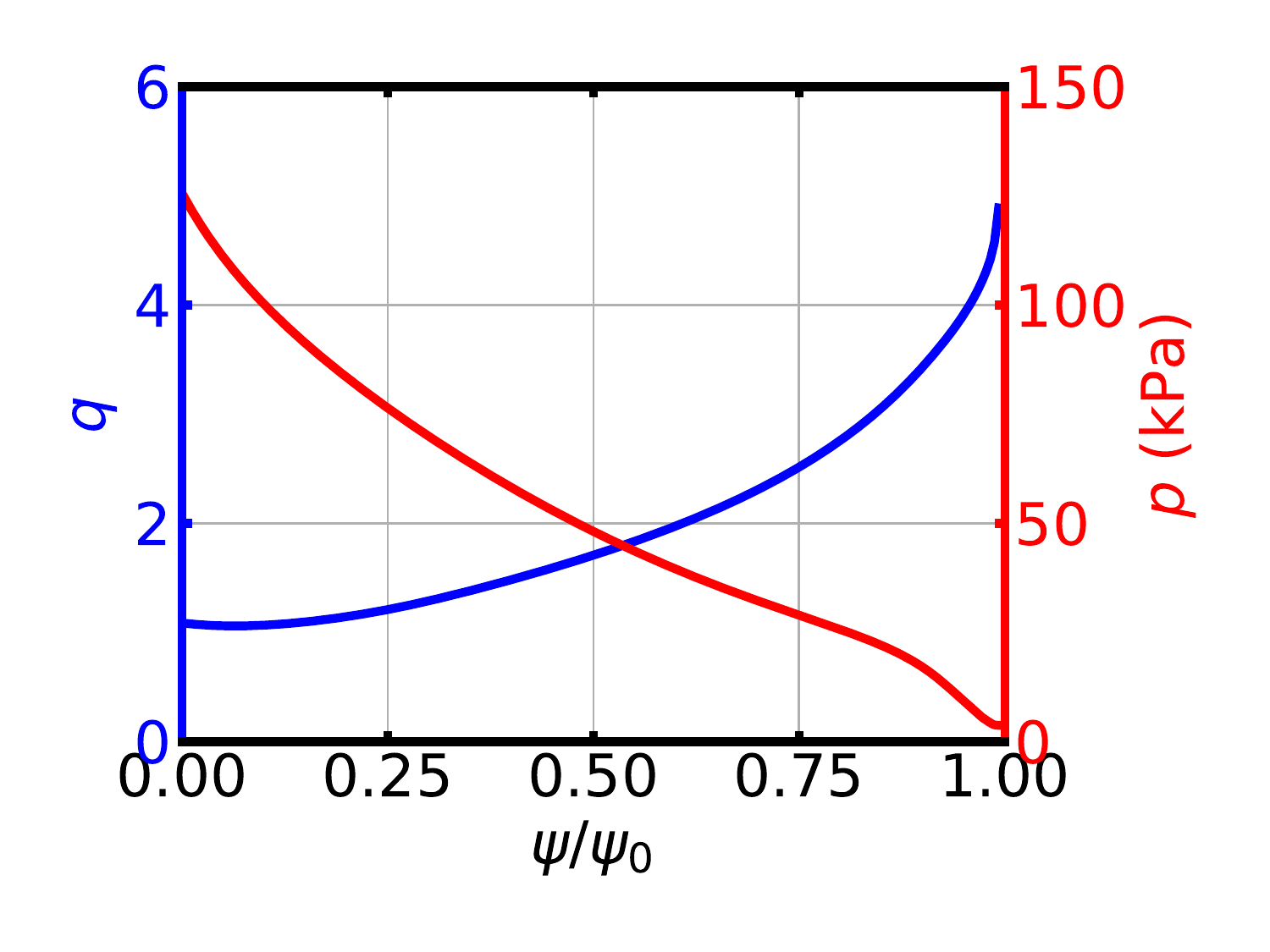}
			\put(70,20) {\textsf{(a)}}
		\end{overpic}
		\raisebox{-0.13\height}{\begin{overpic}[width=0.35\textwidth]{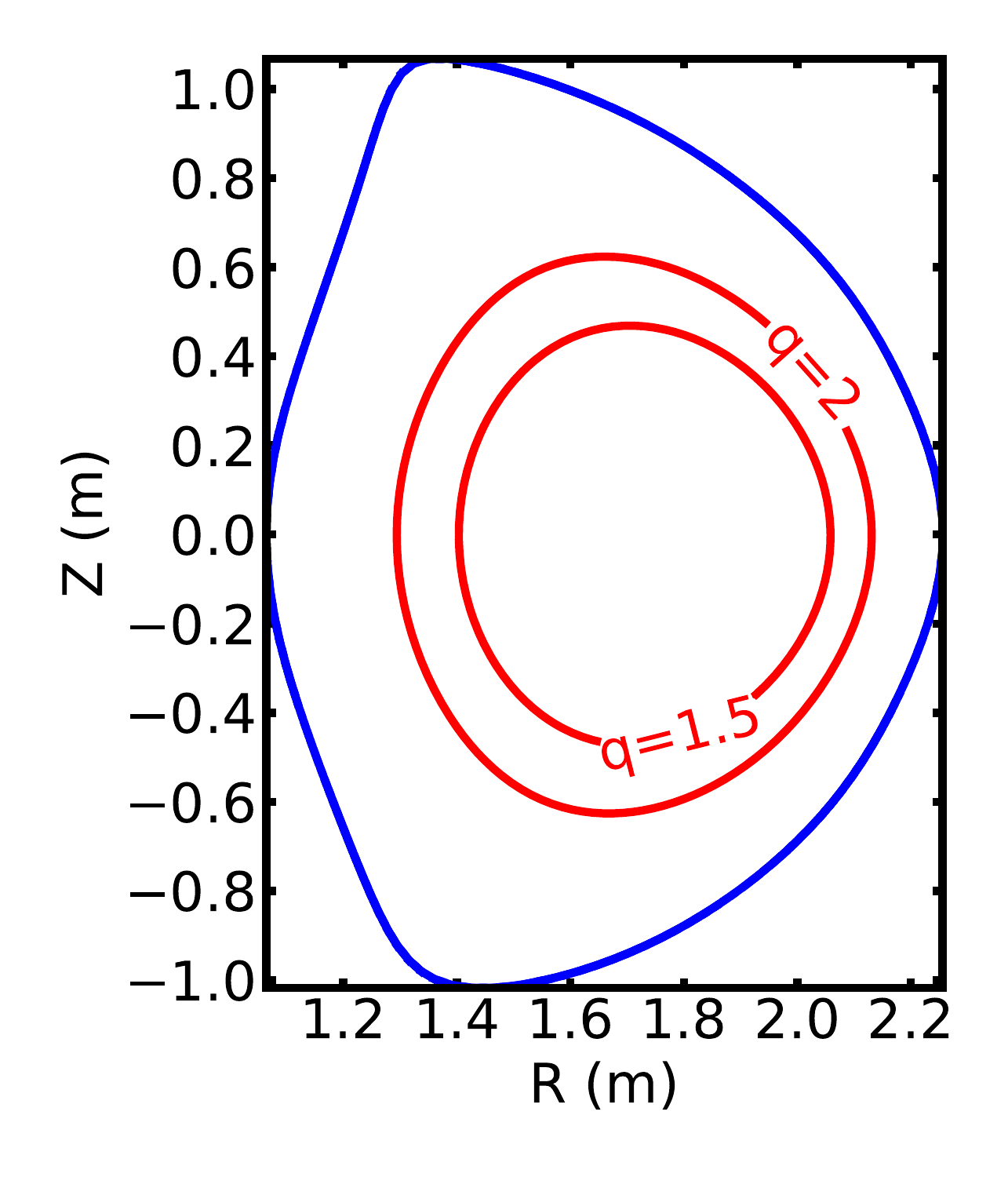}
			\put(70,20) {\textsf{(b)}}
		\end{overpic}}
	\end{center}
	\caption{(a) Profiles of $q$ and total pressure of the equilibrium used in the DIII-D simulation. (b) Flux contours and mesh boundary used in the simulation.}
	\label{fig:q-profile}
\end{figure}

In this section we discuss the linear $n=1$ fishbone mode simulation using M3D-C1-K without and with thermal ion kinetic effects. We use an equilibrium from the DIII-D tokamak experiment, obtained from hybrid discharge \#125476, which has been studied before for $n=1$ \gls{mhd} instabilities using NIMROD\citep{brennan_energetic_2012}. In the experiment, both a (1,1) kink mode and (2,1) and (3,2) tearing modes are present\citep{la_haye_islands_2010}. We use a single equilibrium reconstruction from the EFIT code\citep{lao_equilibrium_1990} including motional Stark effect (MSE) profile data, by choosing a time (3425ms) during the stationary phase of the discharge with benign tearing mode excitation. The toroidal flow is not included in the simulation. The equilibrium profiles of $q$ and total pressure, and the shape of flux contours, are shown in \cref{fig:q-profile}. For this equilibrium the safety factor has a minimum $q_\mathrm{min}=1.06$ located at $\psi/\psi_0=0.0625$, inside which there is a slightly reverse shear near the core. To study the effect of the $q$ profile on the stability of fishbone modes, we apply the Bateman scaling method\citep{bateman_mhd_1978} to the equilibrium, which means that we add a constant value to $F^2$ ($F=RB_\phi$) to change the toroidal field while keeping the pressure and the toroidal current fixed and the Grad-Shafranov equation satisfied.

Both thermal ions and fast ions from neutral beam injection are included in the kinetic-MHD simulation. Both of the populations are deuterium. The thermal ions are initialized with Maxwellian distribution, with density $n_i=n_e-n_f$ ($n_f$ is the fast ion density) and temperature $T_i=T_e$.  The fast ions have a slowing-down distribution in energy with isotropic distribution in pitch angle,
\begin{equation}
	f_0=\frac{n_f\left(\psi\right)}{\mathcal{E}^{3/2}+\mathcal{E}_c^{3/2}},
\end{equation}
where $\mathcal{E}_c=10$keV is the critical energy. $f_0$ has a cutoff energy $\mathcal{E}_\mathrm{max}=50$keV. Both $\mathcal{E}_c$ and $\mathcal{E}_\mathrm{max}$ are constants in the simulation domain. The density of the fast ions $n_f$ has the same profile as the total pressure, so that the fraction of fast ion pressure to the total pressure is fixed (16\%) at different flux surfaces. 

For linear simulations, we use a two-dimensional finite element mesh and a spectral representation of the \gls{mhd} fields in the toroidal direction. The 2D unstructured mesh has 5495 triangular elements which are uniformly distributed. For the kinetic-MHD simulation we use $8\times 10^6$ particle markers, half of which are for the simulation of fast ions and the other half for the thermal ions.

\begin{figure}
	\centerline{\includegraphics[width=0.55\linewidth]{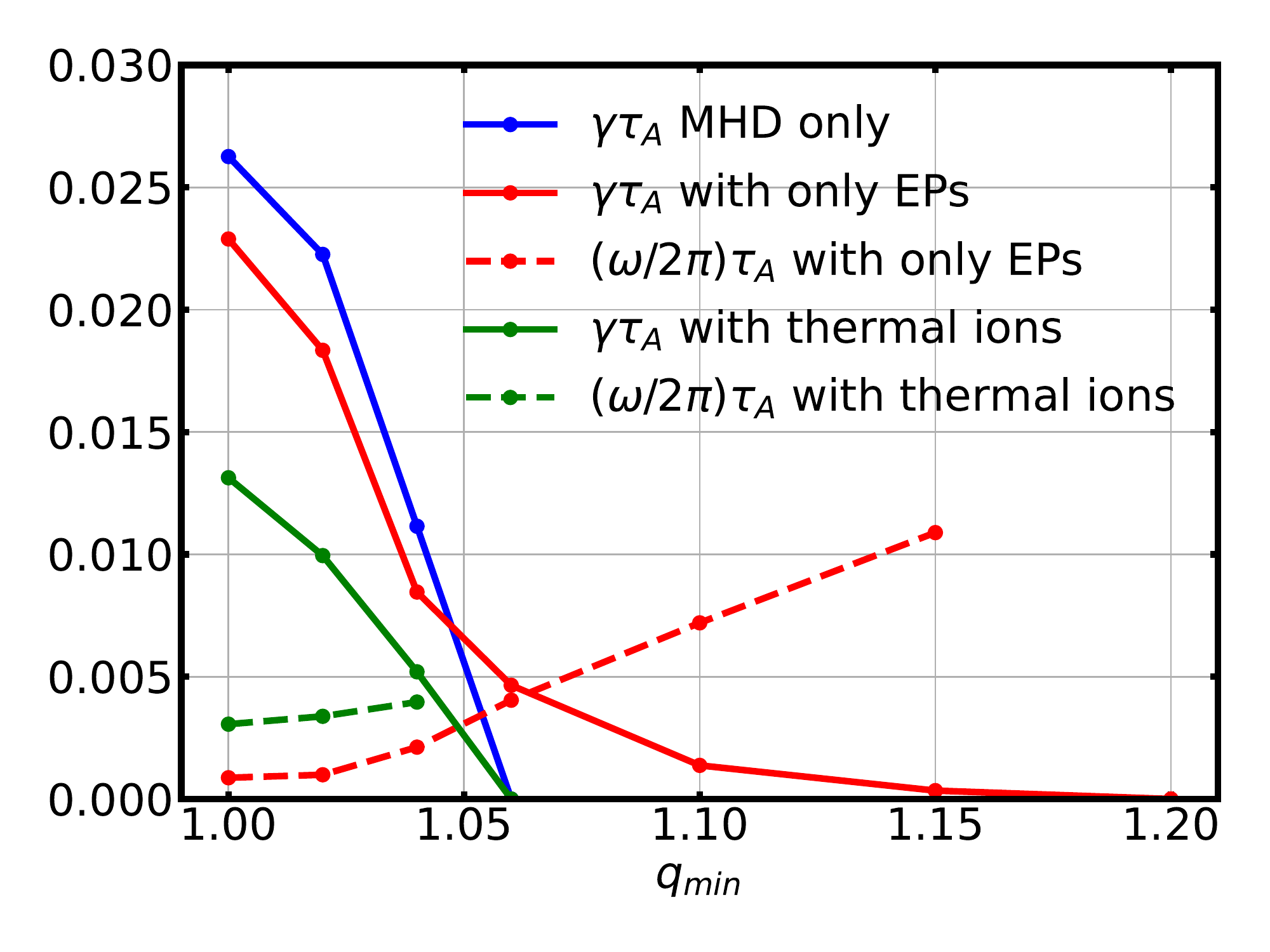}}
	\caption{Growth rates (solid lines) and frequencies (dashed lines) as functions of $q_\mathrm{min}$ of the $n=1$ mode from M3D-C1 linear simulations with DIII-D equilibrium. The blue lines shows the \gls{mhd}-only result. The red lines show the kinetic-MHD results with only fast ions. The green lines show the results with both thermal and energetic ions.}
	\label{fig:d3dkink}
\end{figure}

The simulation results of mode growth rates and frequencies are summarized in \cref{fig:d3dkink}, with $q_\mathrm{min}$ varying from 1.0 to 1.2. The \gls{mhd}-only  result (blue line) shows that the $n=1$ kink mode is unstable for $q_\mathrm{min}<1.06$. After including the fast ion kinetic effects (red lines), the mode growth rates decrease for the $q_\mathrm{min}<1.04$, and the modes have finite frequencies due to the wave-particle resonances, which increase with $q_\mathrm{min}$. For $q_\mathrm{min}>1.1$, there is still a weakly unstable $n=1$ mode which is driven by fast ions, with frequencies significantly larger than those of $q_\mathrm{min}<1.06$ cases. The manifest change of frequencies indicates that the dominant $n=1$ mode changes from a fishbone-like branch to a \gls{bae}-like branch with distinct frequencies. The results of the mode frequencies and growth rates are close to the NIMROD results in \citet{brennan_energetic_2012}, except that the growth rate for  $q_\mathrm{min}>1.1$ keeps dropping to zero as $q_\mathrm{min}$ increases, whereas in \citet{brennan_energetic_2012} the growth rate is almost a constant for large $q_\mathrm{min}$.

\begin{figure}
	\begin{center}
		\begin{overpic}[width=0.33\textwidth]{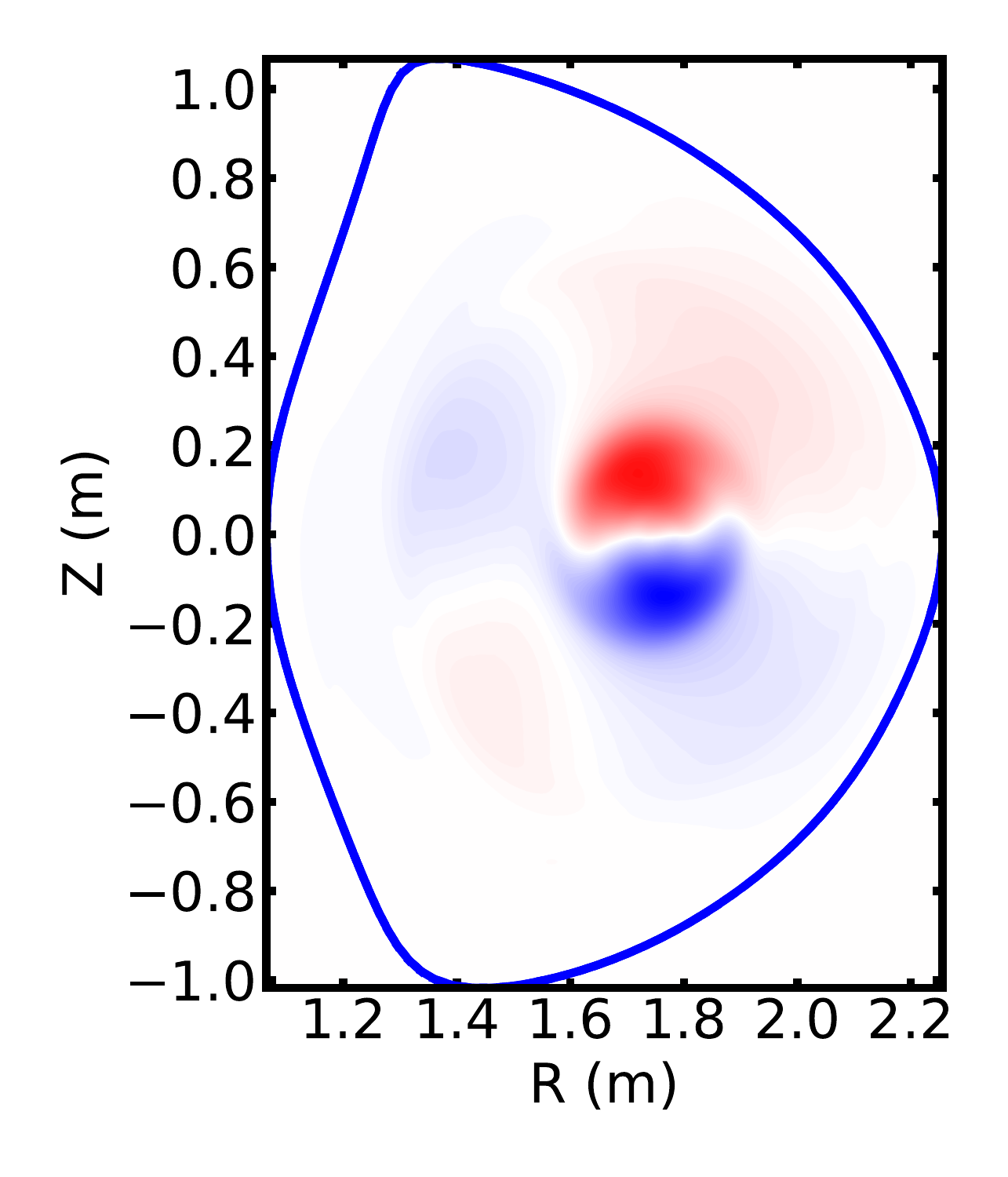}
			\put(60,20) {\textsf{(a)} $\delta \phi$}
		\end{overpic}
		\begin{overpic}[width=0.245\textwidth,trim=95 0 0 0,clip]{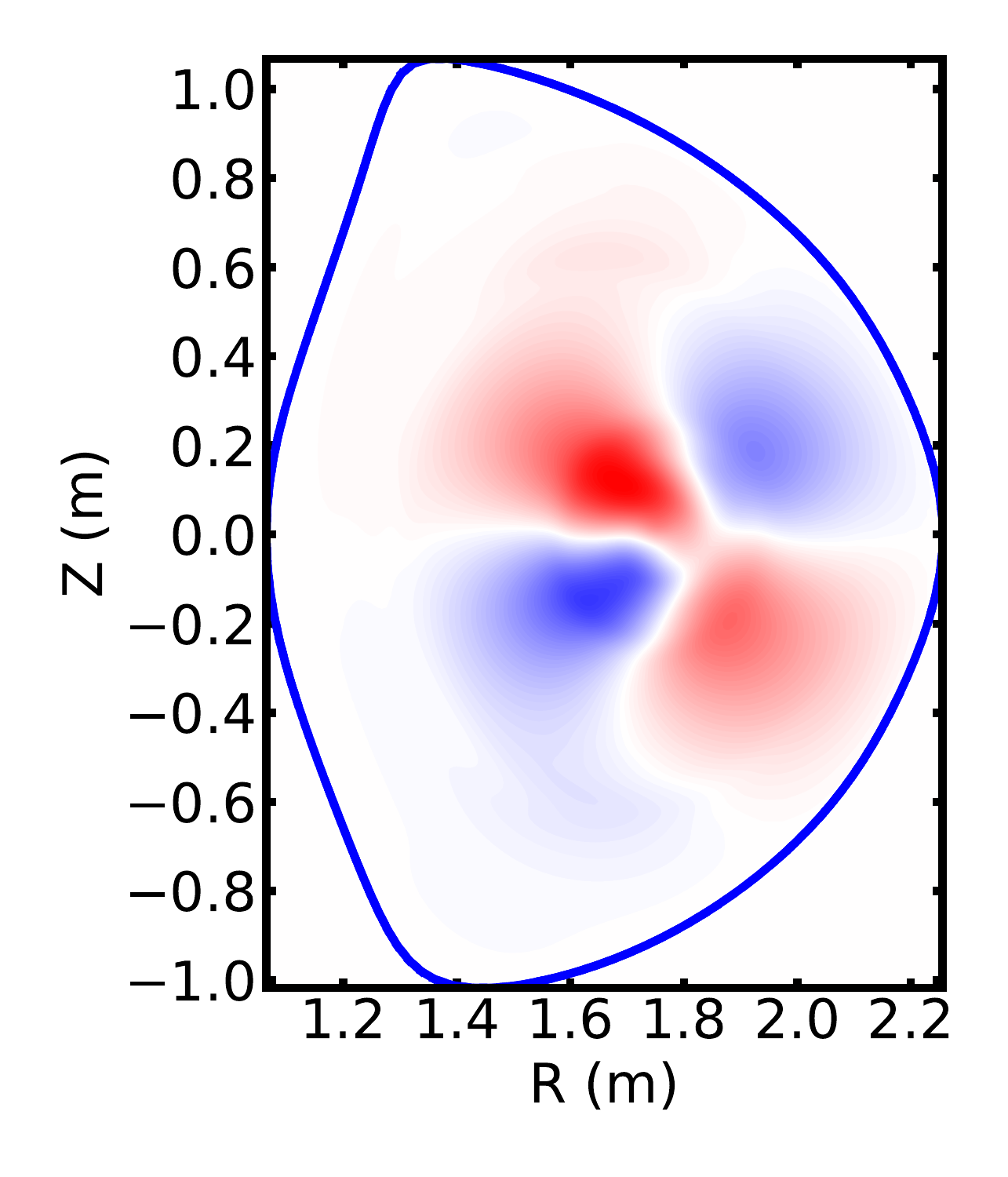}
			\put(38,20) {\textsf{(b)} $\delta \psi$}
		\end{overpic}
		\includegraphics[width=0.077\textwidth,trim=285 0 0 0,clip]{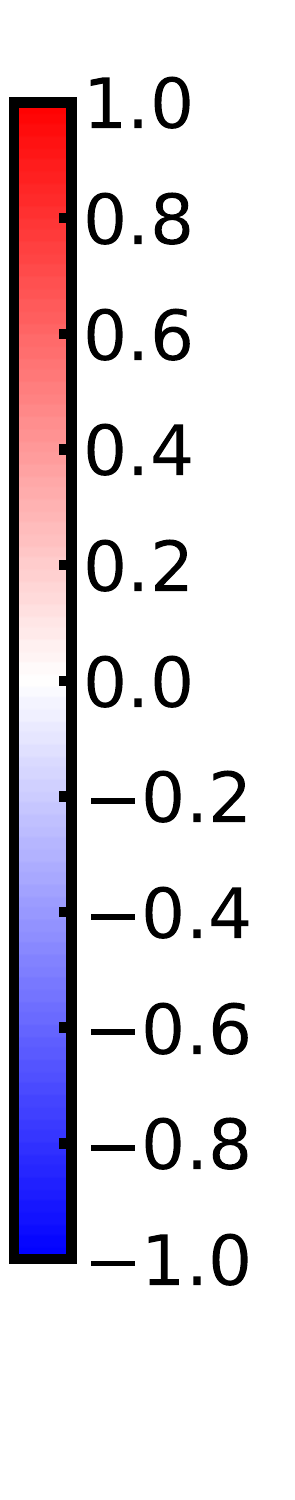}
	\end{center}
	\caption{2D structure of $\delta\phi$ (a) and $\delta\psi$ (b) from DIII-D $n=1$ linear simulation of the $q_\mathrm{min}=1.04$ case with thermal ions.}
	\label{fig:d3dphi}
\end{figure}

\begin{figure}
	\begin{center}
		\begin{overpic}[width=0.33\textwidth]{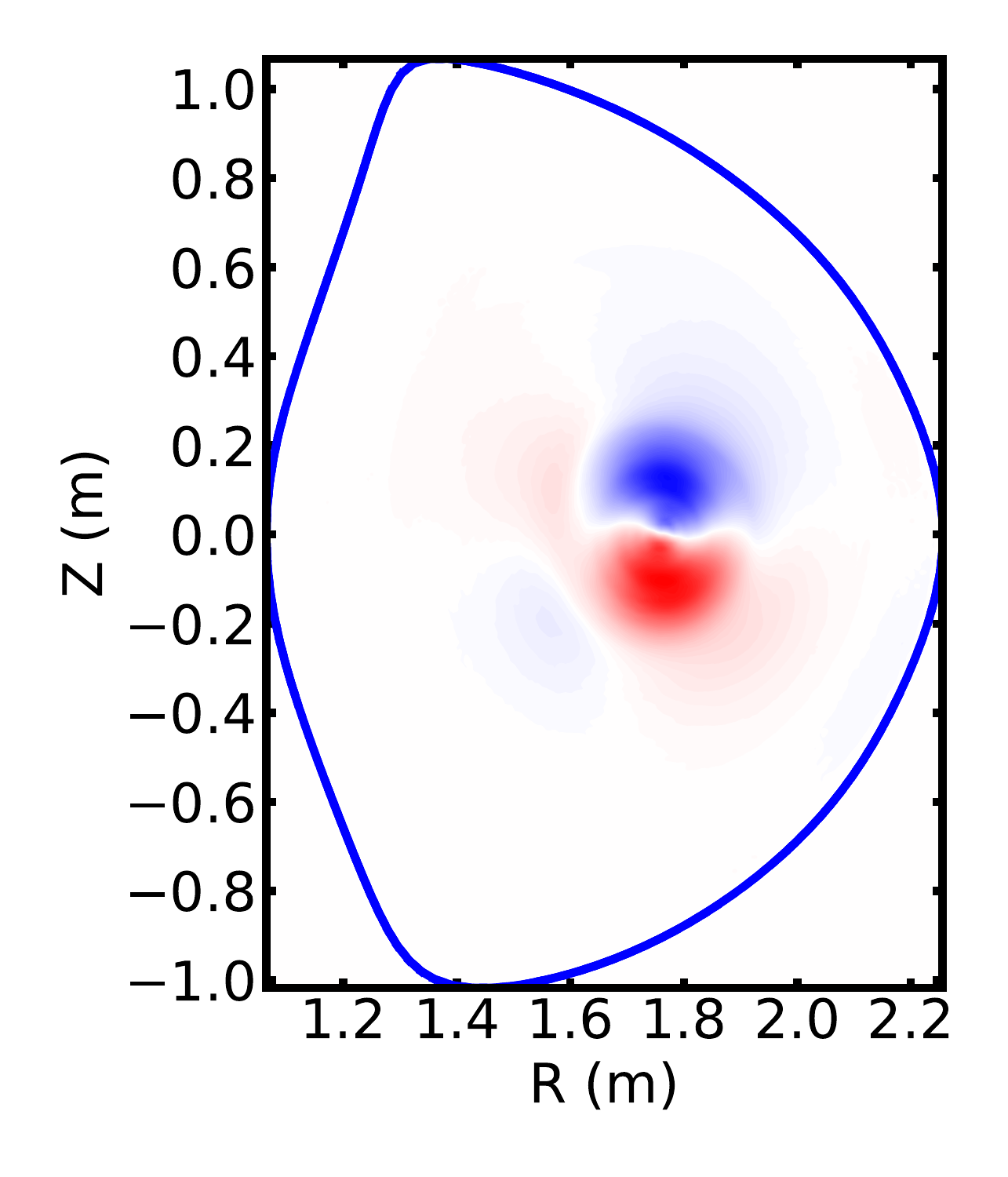}
			\put(60,20) {\textsf{(a)} $\delta p_e$}
		\end{overpic}
		\begin{overpic}[width=0.245\textwidth,trim=95 0 0 0,clip]{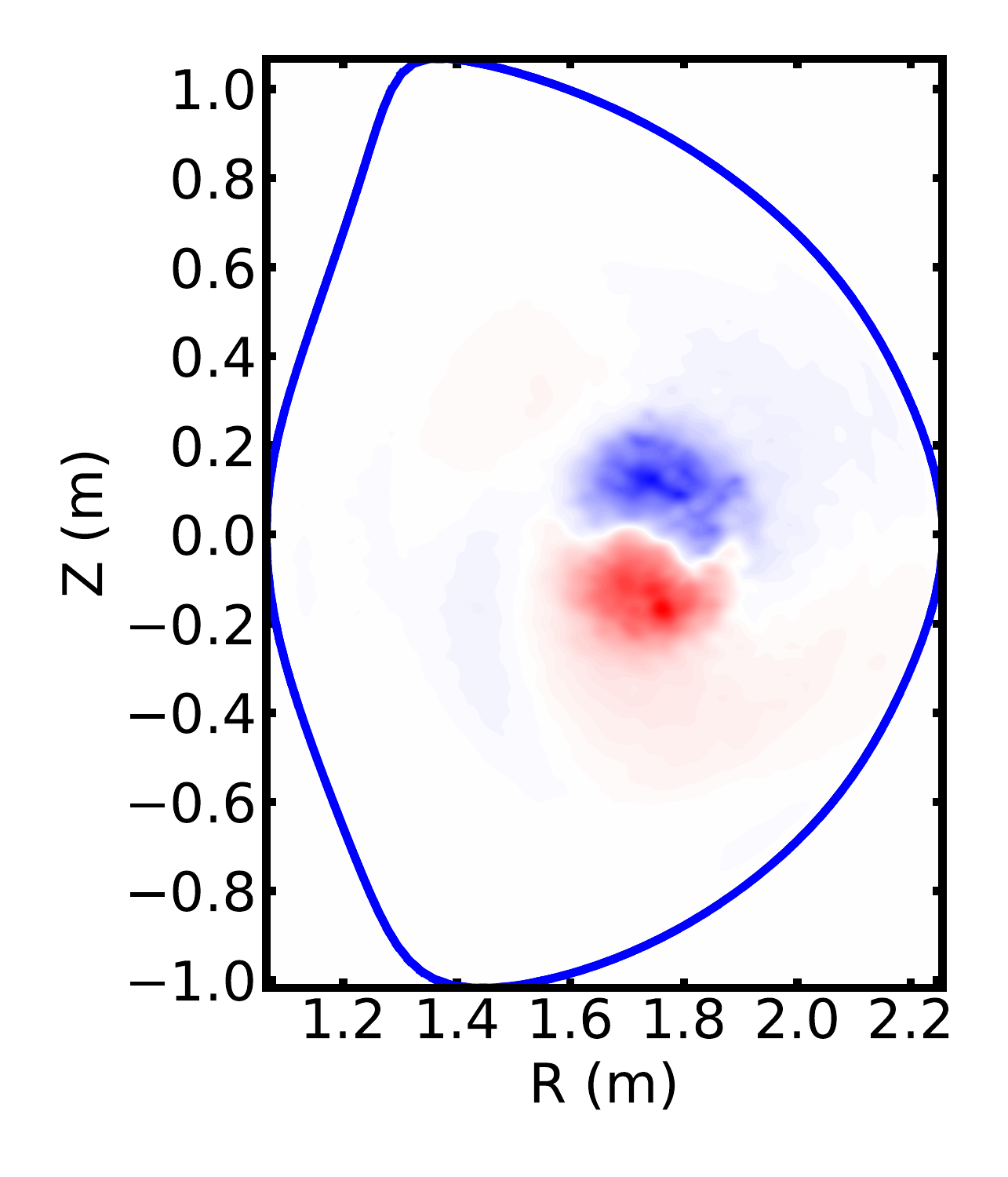}
			\put(38,20) {\textsf{(b)} $\delta p_i$}
		\end{overpic}
		\includegraphics[width=0.077\textwidth,trim=285 0 0 0,clip]{colorbar}
	\end{center}
	\caption{2D structure of perturbed electron pressure $\delta p_e$ (a) and thermal ion pressure $\delta p_i$ (b) from the linear simulation of the $q_\mathrm{min}=1.04$ case with thermal ions.}
	\label{fig:d3dp}
\end{figure}

\begin{figure}
	\begin{center}
		\begin{overpic}[width=0.33\textwidth]{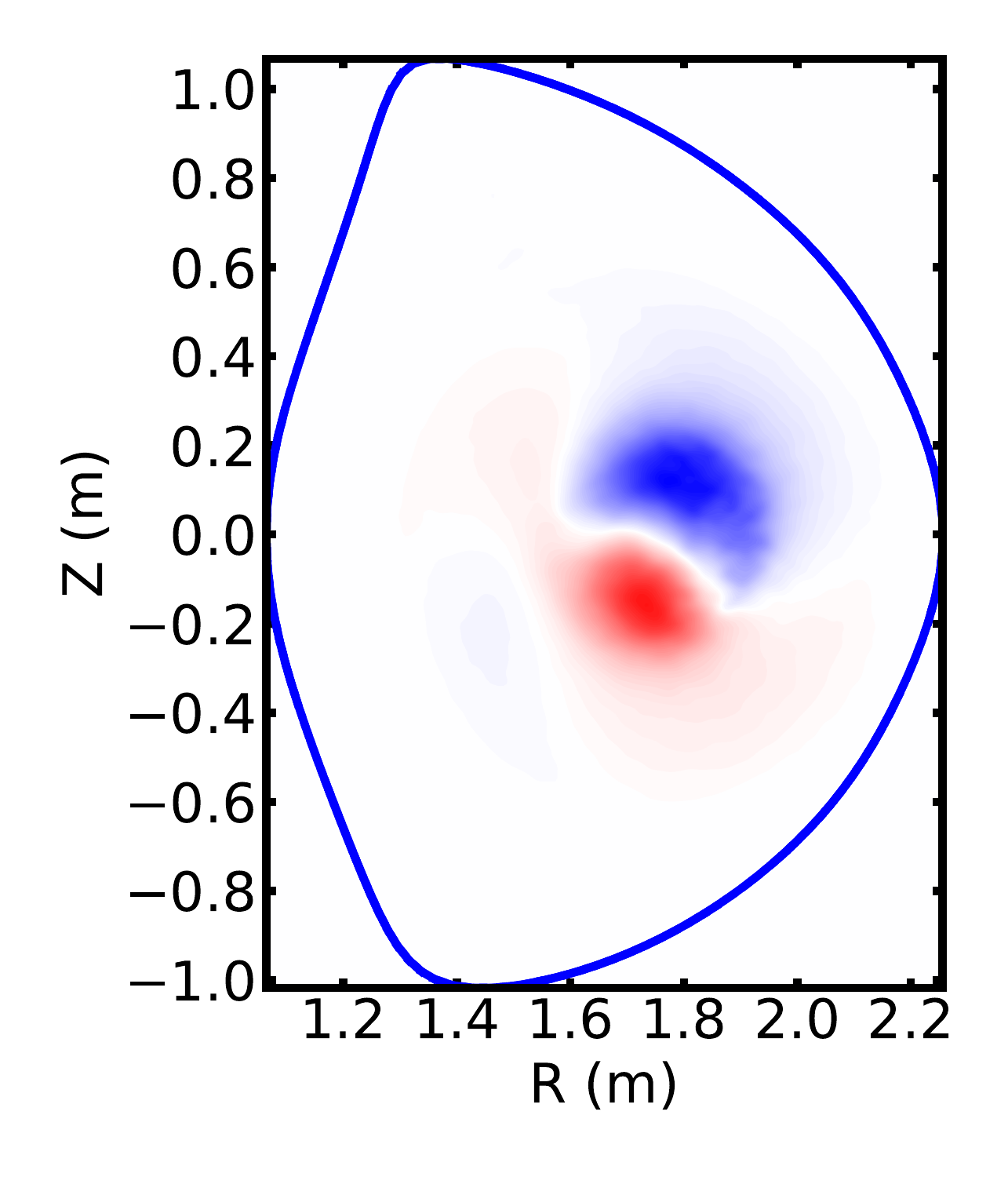}
			\put(50,20) {\textsf{(a)} $\delta p_{\perp,EP}$}
		\end{overpic}
		\begin{overpic}[width=0.245\textwidth,trim=95 0 0 0,clip]{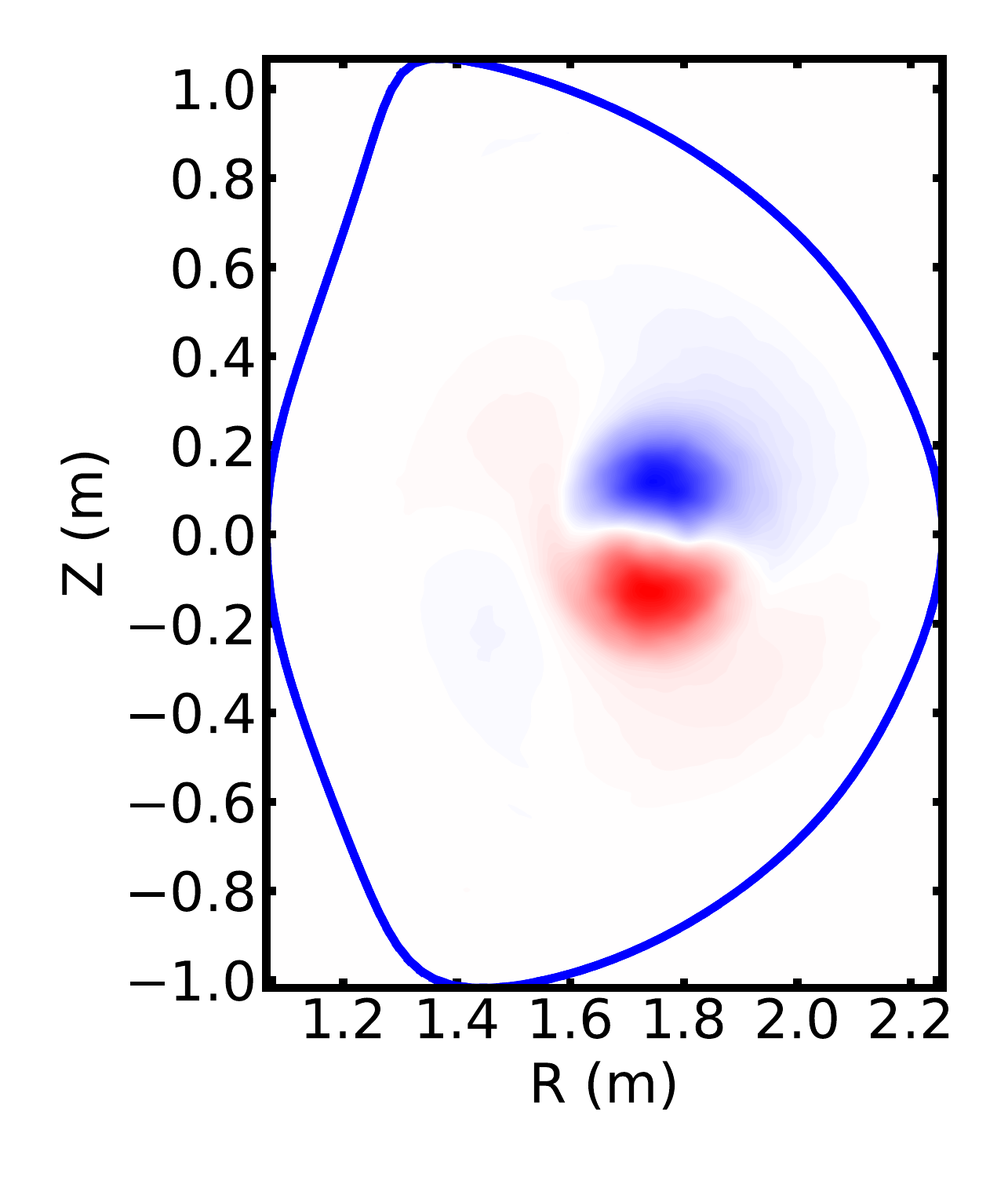}
			\put(29,20) {\textsf{(b)} $\delta p_{\parallel,EP}$}
		\end{overpic}
		\begin{overpic}[width=0.245\textwidth,trim=95 0 0 0,clip]{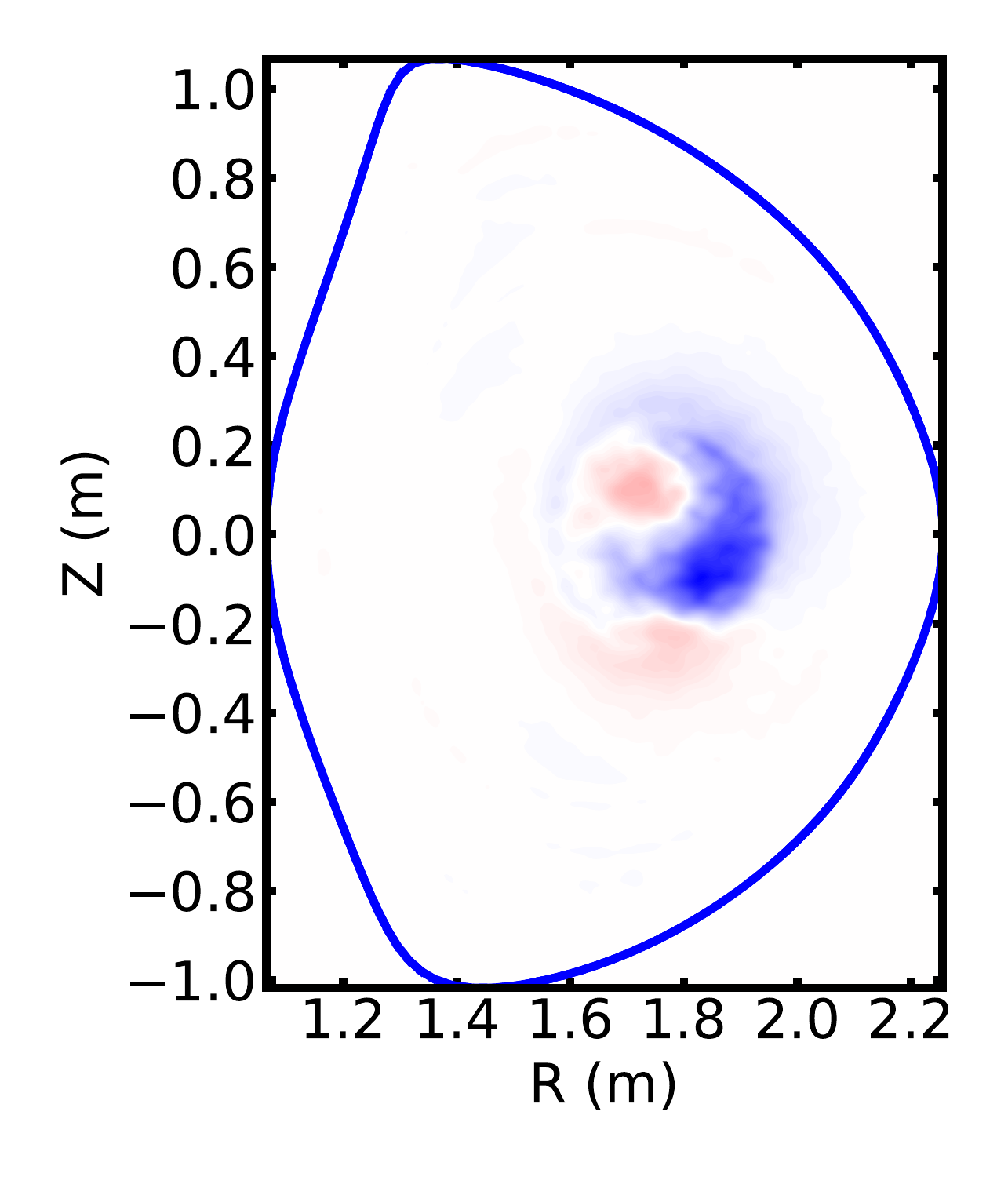}
			\put(2,20) {\textsf{(c)} $\delta p_{\perp,EP}-\delta p_{\parallel,EP}$}
		\end{overpic}
	    \includegraphics[width=0.077\textwidth,trim=285 0 0 0,clip]{colorbar}
	\end{center}
	\caption{2D structure of perpendicular (a) and parallel (b) fast ion pressure from the linear simulation of the $q_\mathrm{min}=1.04$ case with thermal ions, and the difference between the two (c). }
	\label{fig:d3dpf}
\end{figure}

\begin{figure}
	\begin{center}
		\begin{overpic}[width=0.33\textwidth]{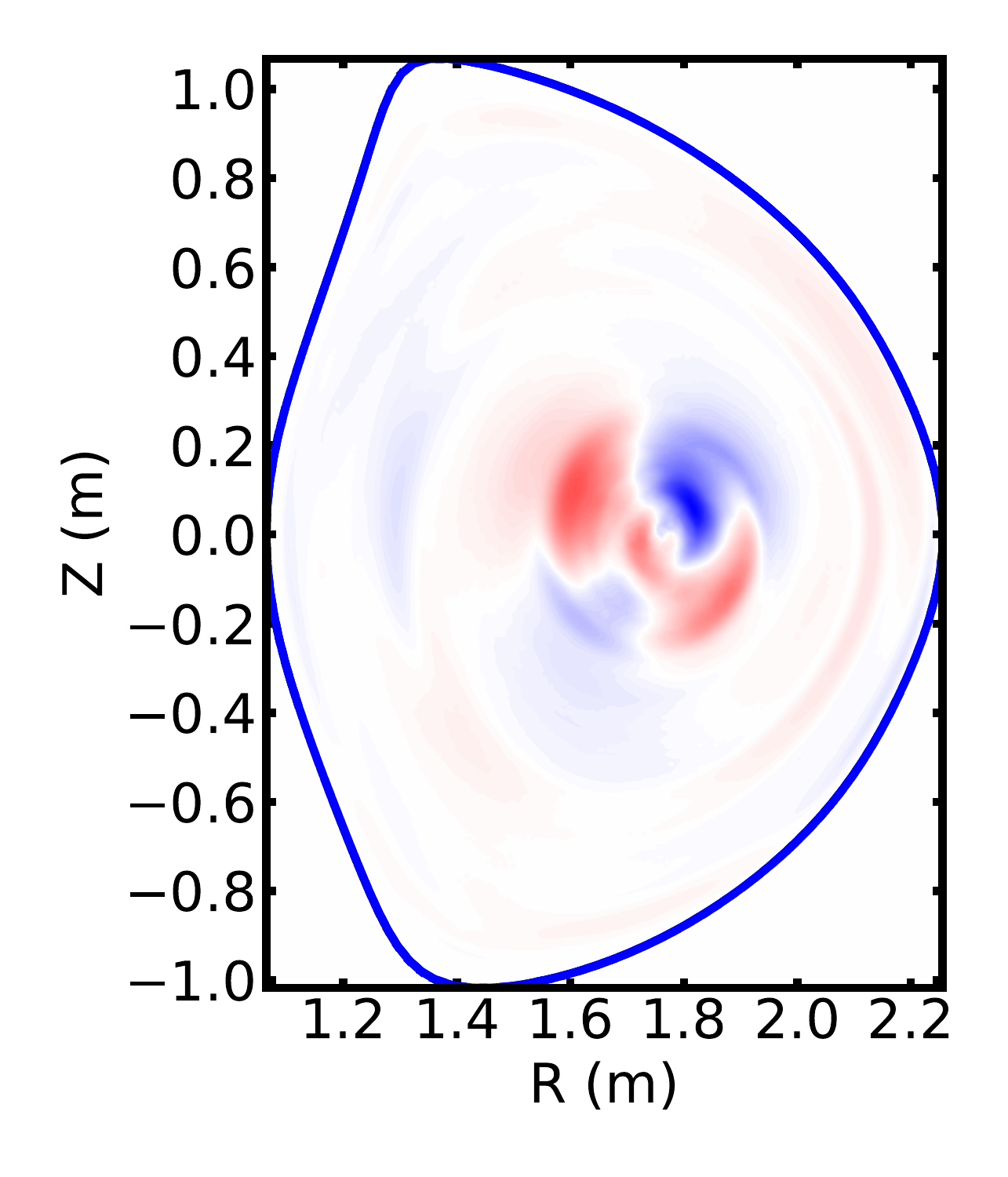}
			\put(42,23) {\parbox{1.7cm}{\textsf{(a) $\delta v_\parallel$ with thermal ions}}}
		\end{overpic}
		\begin{overpic}[width=0.245\textwidth,trim=95 0 0 0,clip]{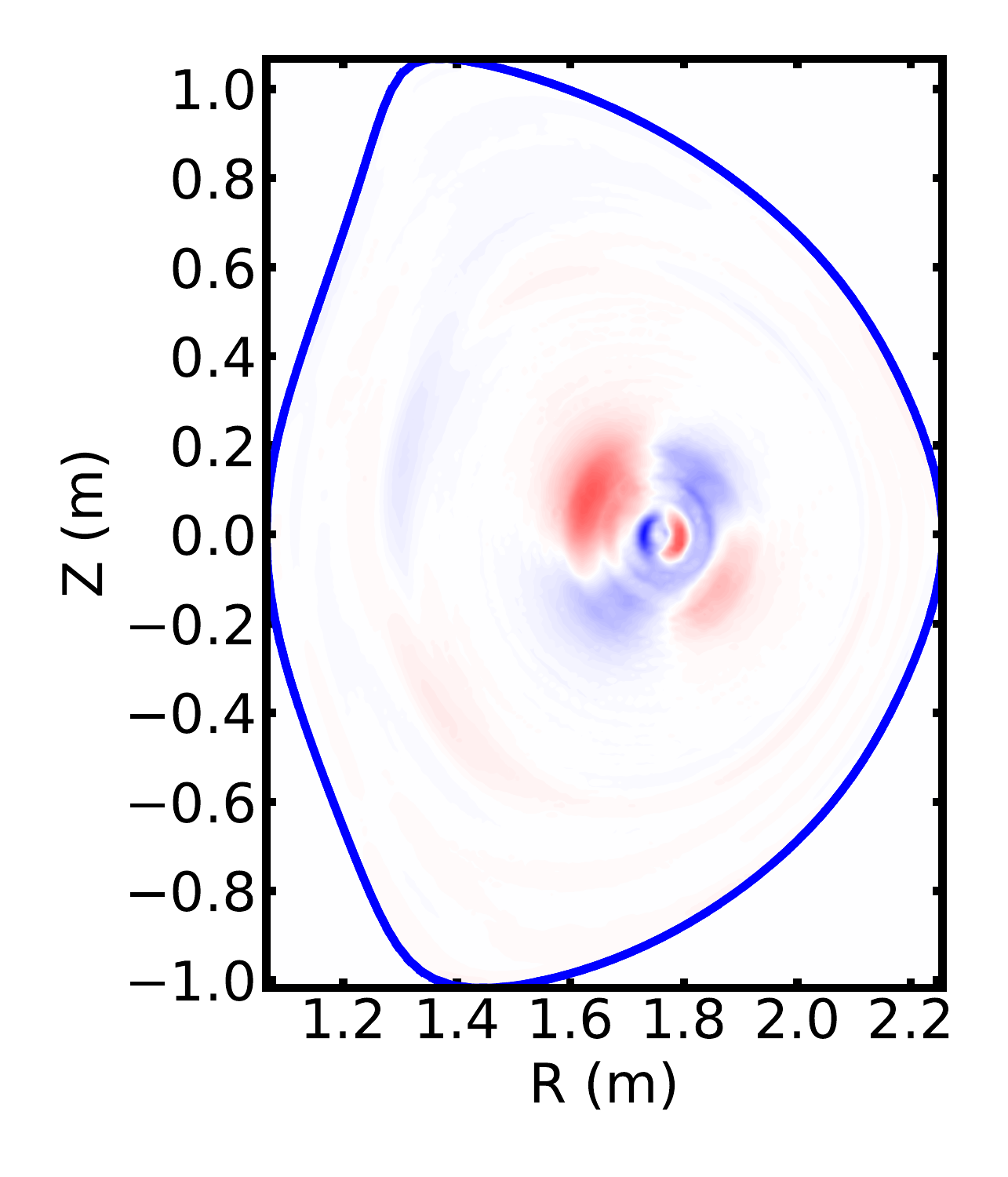}
			\put(22,23) {\parbox{1.7cm}{\textsf{(b) $\delta v_\parallel$ w/o thermal ions}}}
		\end{overpic}
		\includegraphics[width=0.077\textwidth,trim=285 0 0 0,clip]{colorbar}
	\end{center}
	\caption{(a) Structure of $v_\parallel$ from the linear simulation of the $q_\mathrm{min}=1.04$ case with thermal ions and synchronization of $v_\parallel$ (\cref{eq:vpar}). (b) Structure of $v_\parallel$ from the linear simulation with only fast ions using \gls{mhd} equation \cref{eq:dvpardt}.}
	\label{fig:d3dvpar}
\end{figure}

The 2D structure of the fishbone mode for $q_\mathrm{min}=1.04$ including thermal ions is summarized in \cref{fig:d3dphi}-\ref{fig:d3dvpar}. \cref{fig:d3dphi} shows the mode structure of perturbed velocity stream function ($\delta\phi$) and magnetic flux ($\delta\psi$). $\delta\phi$ is dominated by $m=1$ near the core and $m=2$ at the outer region, and $\delta\psi$ is dominated by $m=2$. \cref{fig:d3dp} shows the structure of electron and ion pressure, which is similar to the $\delta\phi$ structure. The similarity between $\delta p_e$ and $\delta p_i$ indicates that the quasi-neutrality condition is satisfied. \cref{fig:d3dpf} shows the structure of fast ion pressure, including parallel and perpendicular components in \cref{eq:ppar,eq:pperp}. The non adiabatic response ($\delta p_\perp-\delta p_\parallel$) is mostly located in the low-field side, as it comes from the resonant trapped particles\citep{fu_global_2006,kim_impact_2008,liu_hybrid_2022}. \cref{fig:d3dvpar} shows the comparison of $\delta v_\parallel$ from the kinetic-MHD simulation using parallel velocity synchronization (\cref{eq:vpar}), with the result of the simulation with only fast ions and no synchronization. The two results are close and both are dominated by $m=2$, showing that the kinetic equations successfully captures the parallel dynamics of the \gls{mhd} system.

According to the simulation results, we find that both the fishbone branch and the \gls{bae}-like branch of the $n=1$ mode are susceptible to Landau damping of thermal ions. The parallel wave number can be estimated as follows,
\begin{equation}
	k_\parallel\approx\frac{1}{R}\left(n-\frac{m}{q_\mathrm{min}}\right).
\end{equation}
Here we choose $m=1.5$ as the mode has both $m=1$ and $m=2$ components. Using the mode frequency obtained from simulation with fast ions, for $q_\mathrm{min}=1.04$ there is $\omega/k_\parallel=0.20 v_\mathrm{th}$, and for $q_\mathrm{min}=1.15$, there is $\omega/k_\parallel=1.15 v_\mathrm{th}$, where $v_\mathrm{th}=\sqrt{T_i/m_i}$. Therefore the Landau damping effects are important for both branches, which explains the stabilization of the mode by thermal ions for $q_\mathrm{min}>1.06$ cases providing their small growth rates without thermal ions.  It is necessary to include the thermal ion kinetic effects and Landau damping in those case to avoid false positive results. 

\begin{figure}
	\centerline{\includegraphics[width=0.55\linewidth]{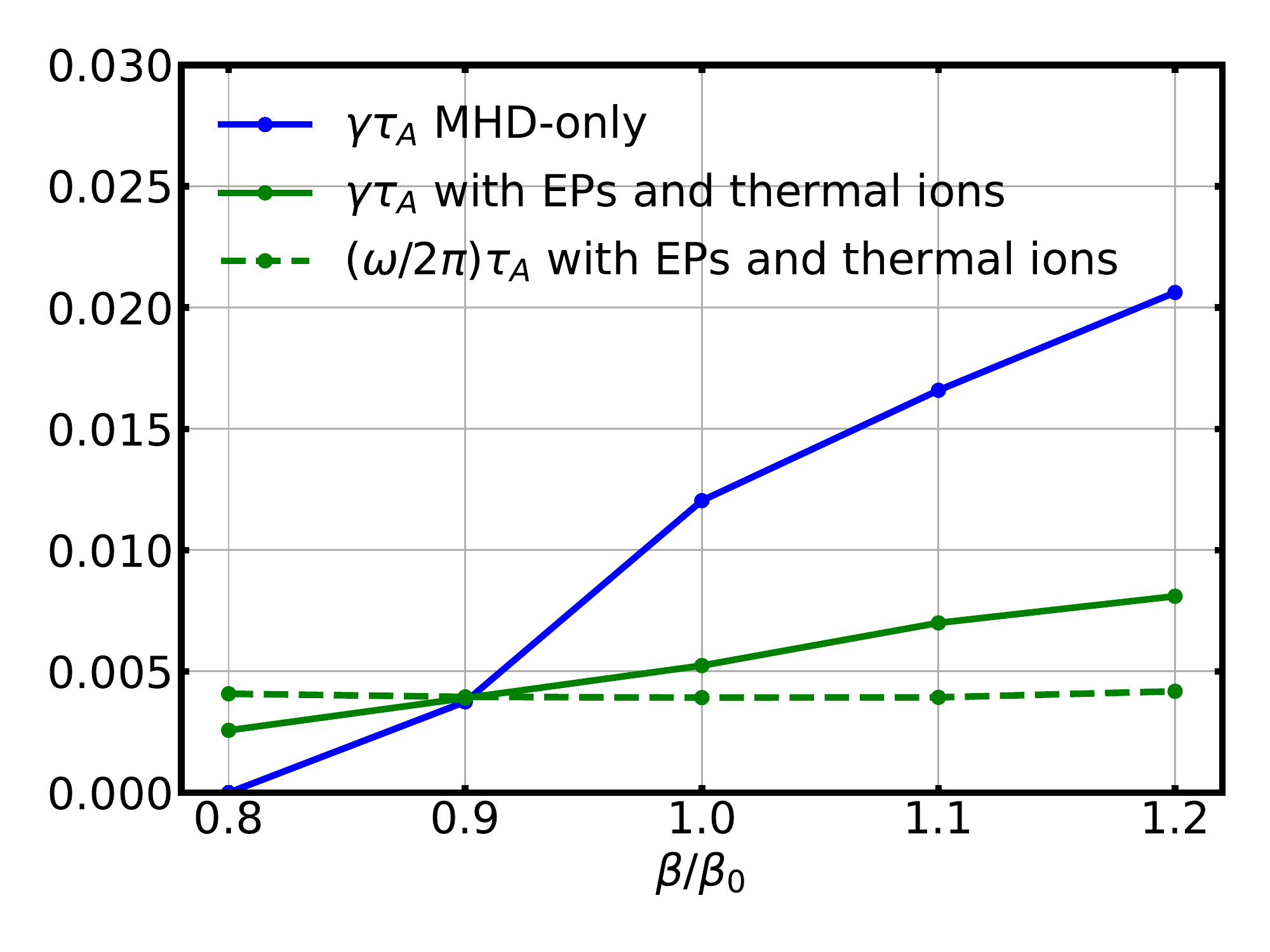}}
	\caption{Growth rates (solid lines) and frequencies (dashed lines) of the $n=1$ mode with different plasma $\beta$ ($\beta_0$ is the experimental value) and a fixed $q$ profile. The blue lines shows the \gls{mhd}-only result. The green lines show the kinetic-MHD results with fast and thermal ions.}
	\label{fig:d3dkink-pressure}
\end{figure}

In the above study we scan the value of $q_\mathrm{min}$ by varying the toroidal field and keep the plasma pressure fixed. This can leads to change to both $q_\mathrm{min}$ and plasma $\beta$ at the same time, as discussed in \citet{brennan_energetic_2012}. To separate the two effects, we rerun the reconstruction of the equilibrium by fixing the toroidal field and $q$ profile while scaling the total pressure, by varying $T_e$, $T_i$ and fast ion energy. The results of MHD-only simulations and simulations with fast ion and thermal ion kinetic effects are summarized in \cref{fig:d3dkink-pressure}, with $q_\mathrm{min}=1.04$. The results indicate that, although the (1,1) mode bears the name "non-resonant kink (NRK) mode" in some literature\citep{wang_linear_2013}, is growth rate has a strong dependence on the plasma beta value, especially for the MHD-only simulations, which is similar to pressure-driven modes. After including ion kinetic effects, the dependence of growth rate on $\beta$ weakens, and the mode frequency is almost independent of $\beta$. 

With the Landau damping effect included, the $n=1$ mode is stable for $q_\mathrm{min}>1.06$. For the original equilibrium from EFIT, $q_\mathrm{min}=1.06$, the mode is at the stability boundary, which may be the result of nonlinear evolution of fishbone mode and the flattening of the current profile near the core. The fishbone-like mode in the simulation with $q_\mathrm{min}=1.04$ has frequency about $f=6.13$ kHz. In the DIII-D experiment, the dominant $n=1$ mode is identified to have a frequency of 18 kHz in the lab frame, and the toroidal rotate frequency is about 21 kHz. Therefore the simulation of fishbone-like mode is consistent with the measured frequency in the plasma frame.

\section{Linear simulation of fishbone modes in NSTX}
\label{sec:nstx-fishbone}

\begin{figure}
	\begin{center}
		\begin{overpic}[width=0.55\textwidth]{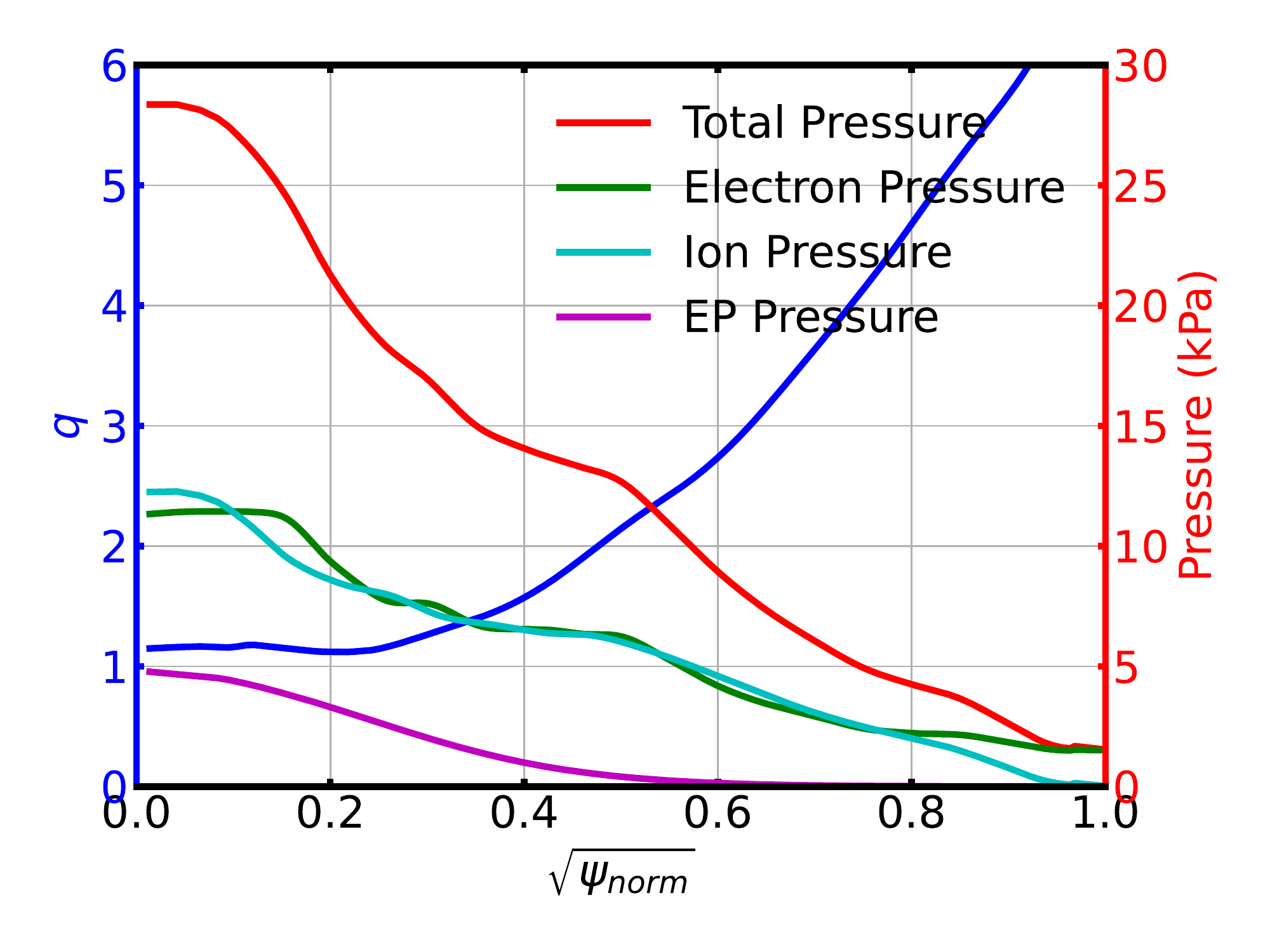}
			\put(80,15) {\textsf{(a)}}
		\end{overpic}
		\raisebox{-0.07\height}{\begin{overpic}[width=0.35\textwidth]{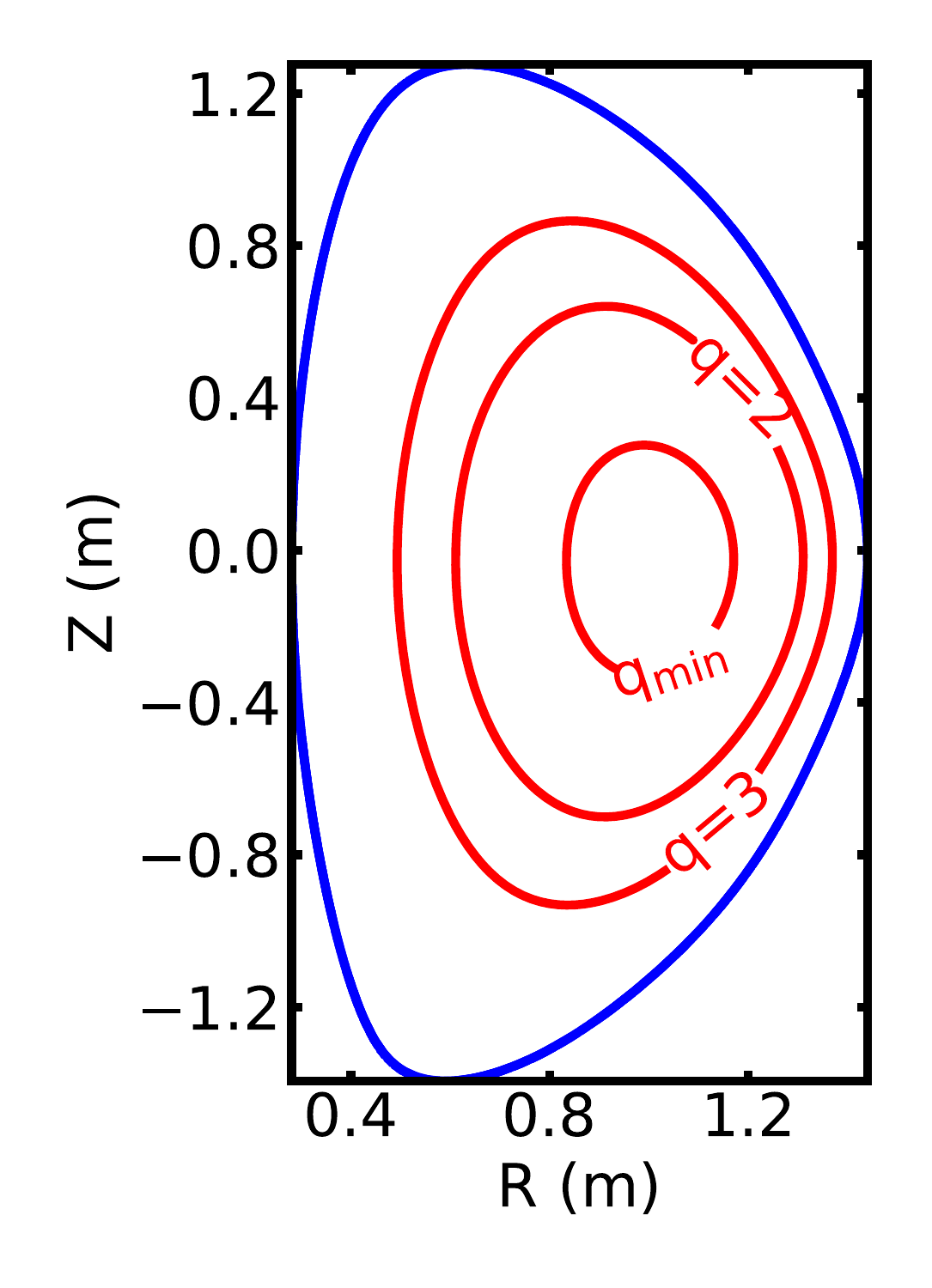}
				\put(60,20) {\textsf{(b)}}
		\end{overpic}}
	\end{center}
	\caption{(a) Profiles of $q$ and pressure of different particle species of the equilibrium used in the NSTX simulation. (b) Flux contours and mesh boundary used in the simulation.}
	\label{fig:q-profile-nstx}
\end{figure}

We did a similar study for the $n=1$ mode under a NSTX experimental condition. The equilibrium $q$ profile is obtained from NSTX shot \#134020 at 700ms. The density and temperature profiles are from the TRANSP\citep{ongena_numerical_2012} calculation result. The \gls{gs} equation was solved using these profiles in a M3D-C1 2D mesh with 7199 elements. We ignored the contribution of high-$Z$ impurities and assume all the ions are deuterium. The profiles used in the simulation and the shape of flux contours are shown in \cref{fig:q-profile-nstx}. The core electron density is 1.04$\times 10^{20}$m$^{-3}$. and the core electron and ion temperature is about 0.74 keV.  The ratio of \gls{ep} beta to the total beta at core is $\beta_\mathrm{EP}/\beta=17.3\%$. Note that a key difference from the DIII-D equilibrium is that NSTX has much large plasma $\beta$ ($\beta_\mathrm{on-axis}=50.8\%$ vs. DIII-D $\beta_\mathrm{on-axis}=12.4\%$).  In the NSTX simulation we use Spitzer resistivity calculated from local electron temperature, and apply a large value of parallel heat conduction coefficient $\kappa_\parallel$.

\begin{figure}
	\begin{center}
		\begin{overpic}[width=0.45\textwidth]{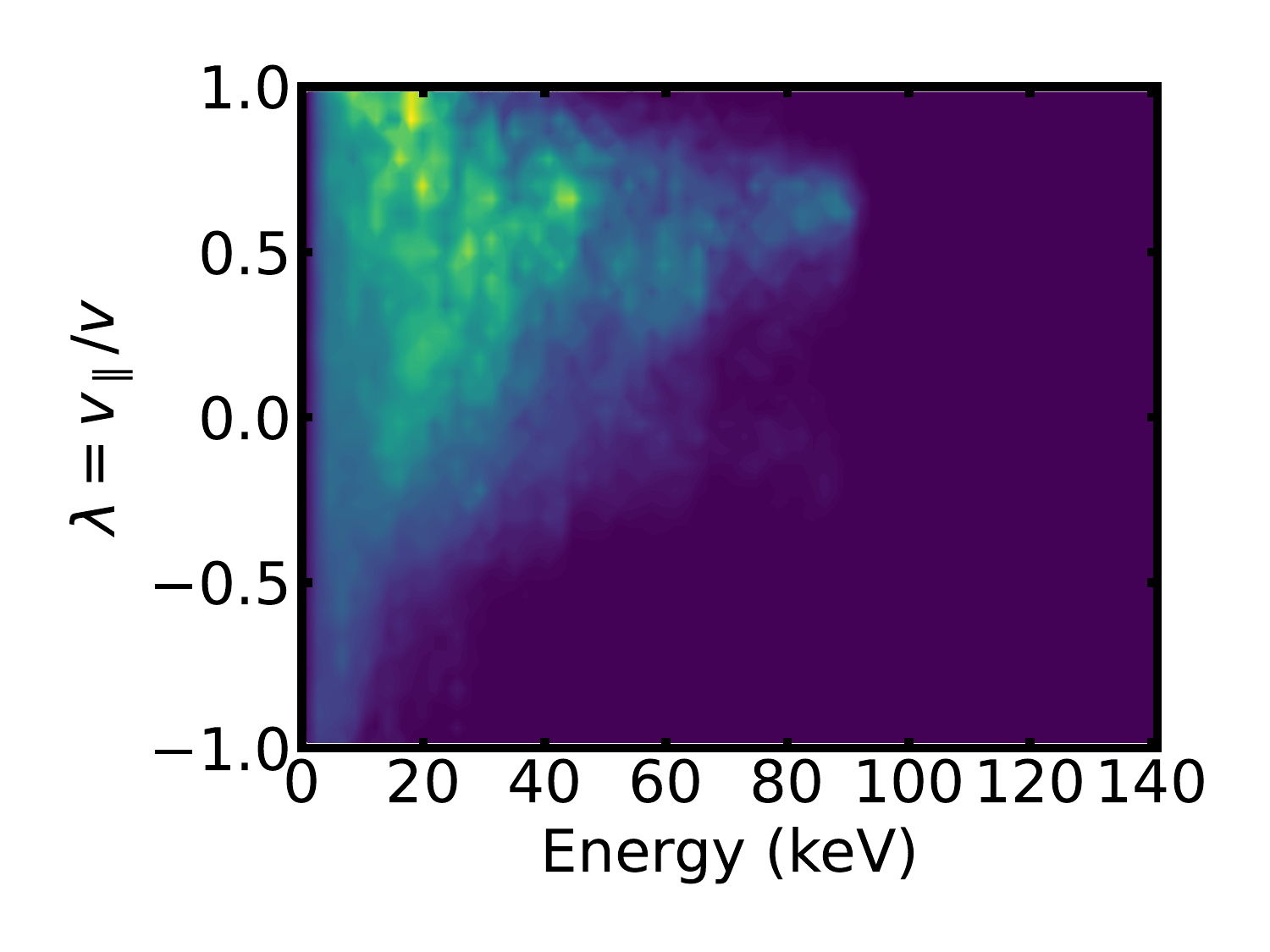}
			\put(83,20) {\color{white}\textsf{(a)}}
		\end{overpic}
		\begin{overpic}[width=0.35\textwidth,trim=97 0 0 0,clip]{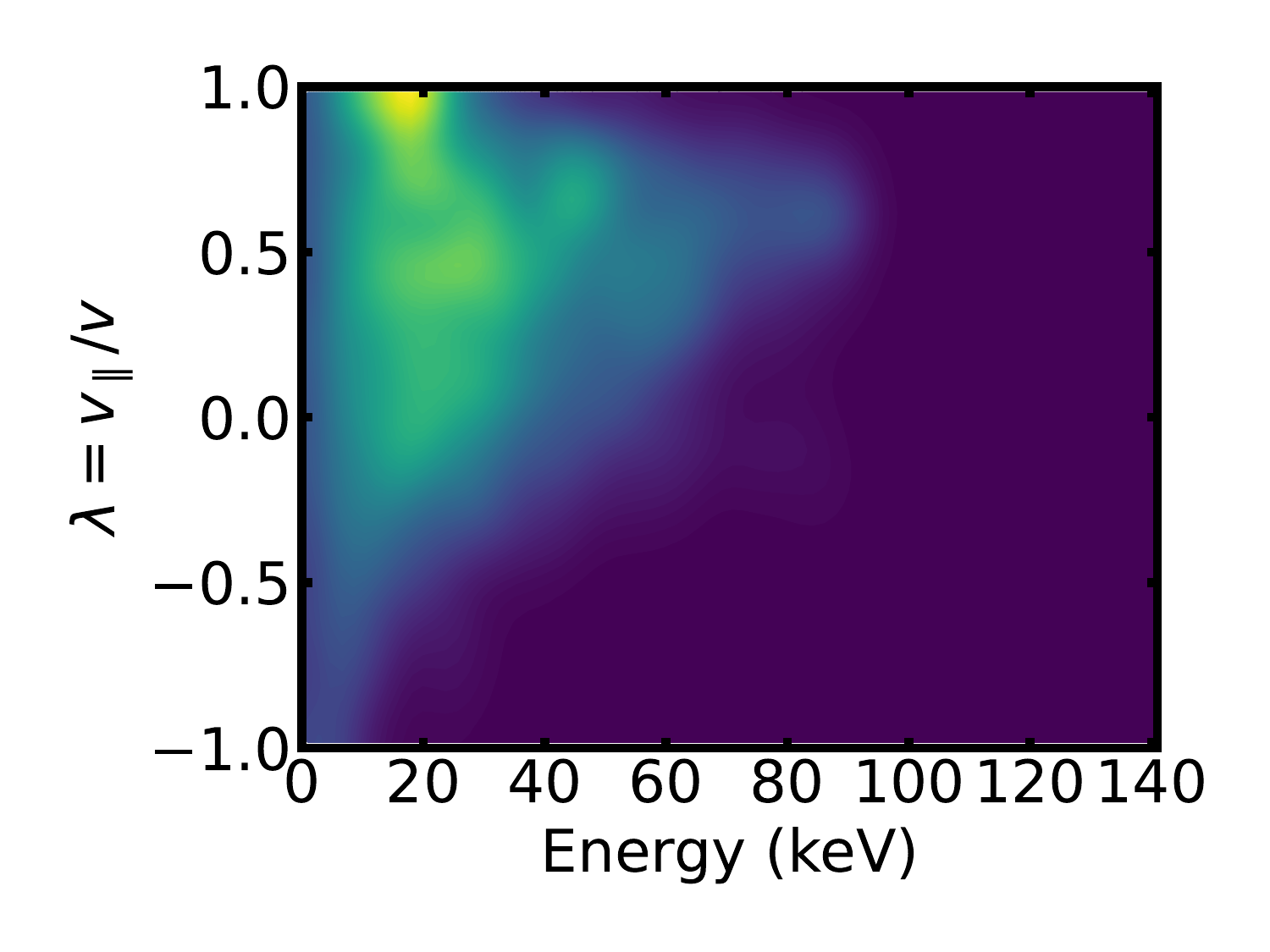}
				\put(77,25) {\color{white}\textsf{(b)}}
		\end{overpic}
	    \raisebox{0.11\height}{\includegraphics[width=0.07\textwidth,trim=355 0 0 0,clip]{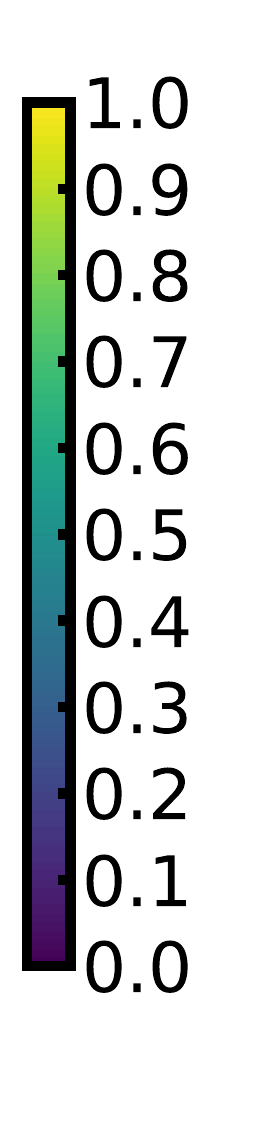}}
	\end{center}
	\caption{(a) The original \gls{ep} distribution of energy and pitch angle near the magnetic axis from NUBEAM. (b) The smoothed \gls{ep} distribution that was used in M3D-C1-K simulations.}
	\label{fig:f-contour}
\end{figure}

For \gls{ep} we use an anisotropic distribution from the NUBEAM calculation. The NUBEAM code provides 3D ($r$, $\lambda=V_\parallel/V$, energy) information of injected beam ions from Monte Carlo calculation\citep{pankin_tokamak_2004}, which is called the classical fast ion distribution. The NUBEAM distribution near the magnetic axis (\cref{fig:f-contour} (a)) shows that the lower energy \glspl{ep} ($\mathcal{E}<40$keV) are mostly co-passing with $V_\parallel/V\approx 1$, and the higher energy \glspl{ep} have a peak distribution at $V_\parallel/V\approx 0.4$. This initial distribution is quite noisy, making it difficult to calculate the gradients of $f_0$ in the phase space. In order to use it for $\delta f$ calculation, we apply a Gaussian smoothing operator to obtain a smoothed distribution as shown in \cref{fig:f-contour} (b). This new distribution is then read into M3D-C1-K as $f_0$ and used for both particle initialization and $\delta f$ calculation. The radial profile of \gls{ep} pressure is shown in \cref{fig:q-profile-nstx}, which is consistent with the TRANSP output.

\begin{figure}
	\centerline{\includegraphics[width=0.55\linewidth]{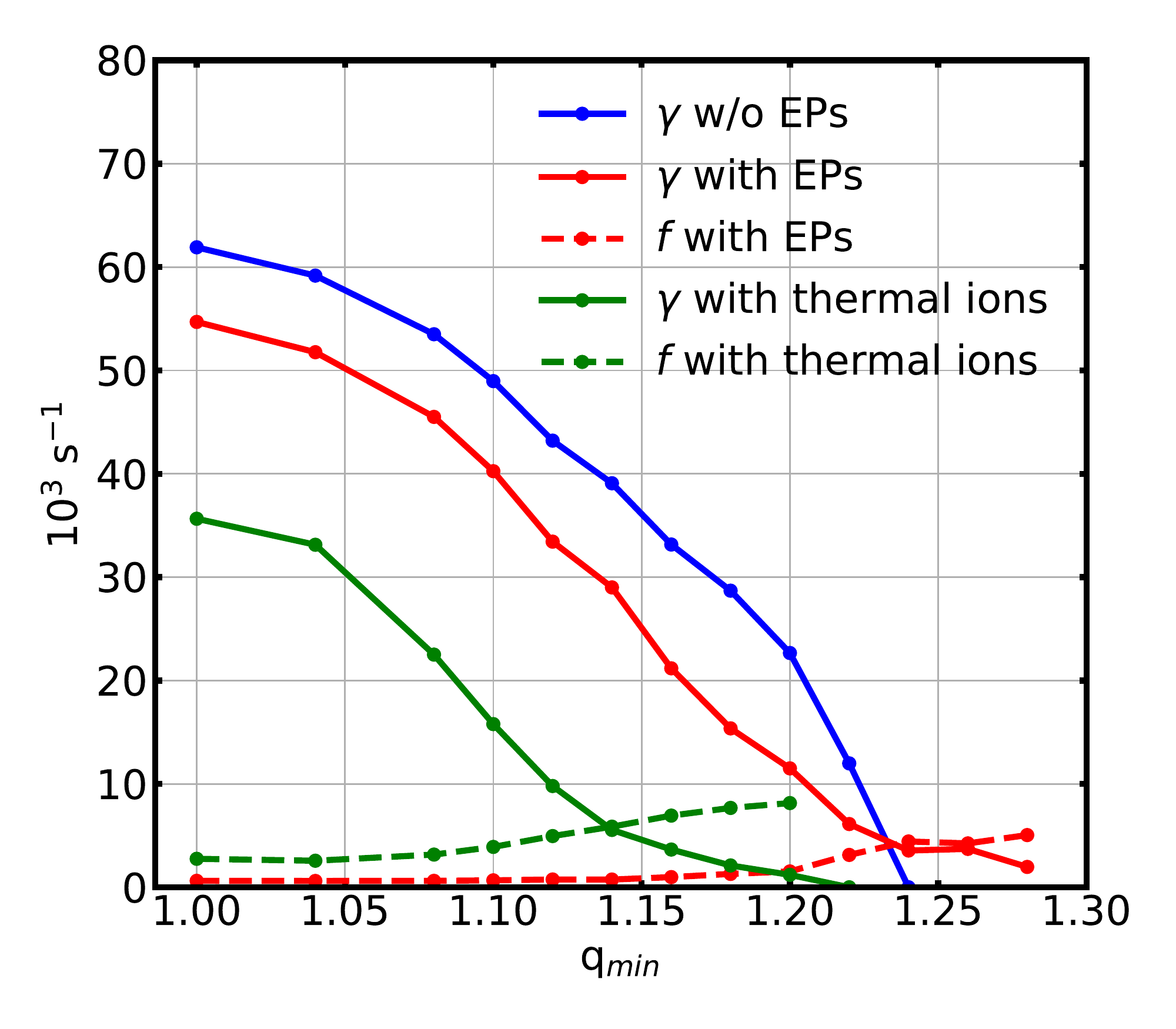}}
	\caption{Growth rates $\gamma$ (solid lines) and frequencies $f$ (dashed lines, in kHz) as functions of $q_\mathrm{min}$ of the $n=1$ modes from M3D-C1 linear simulation with NSTX equilibrium with fixed $\beta$. The blue line shows the \gls{mhd}-only result. The red lines show the kinetic-MHD results with only fast ions. The green lines show the results with both thermal and energetic ions.}
	\label{fig:nstxkink}
\end{figure}

In terms of the shortcomings of Bateman scaling method, and the sensitive dependence of $n=1$ mode growth rate on $\beta$ as discussed in \cref{sec:d3d-fishbone}, for NSTX we apply a new method to scan the value of $q_\mathrm{min}$, by fixing the toroidal field and changing the plasma current to fit the new $q$ profile. This method ensures that $\beta$ is almost fixed when scanning $q_\mathrm{min}$. The results of kinetic-MHD and MHD-only $n=1$ linear simulations are summarized in \cref{fig:nstxkink}. The threshold value of $q_\mathrm{min}$ for $n=1$ mode excitation is higher than the DIII-D results, thanks to the larger $\beta$. For kinetic-MHD simulations with only fast ions (red lines), the dominant mode changes from fishbone-like to BAE-like as $q_\mathrm{min}$ increases above 1.2, with an increase of mode frequencies.

\begin{figure}
	\begin{center}
		\begin{overpic}[width=0.33\textwidth]{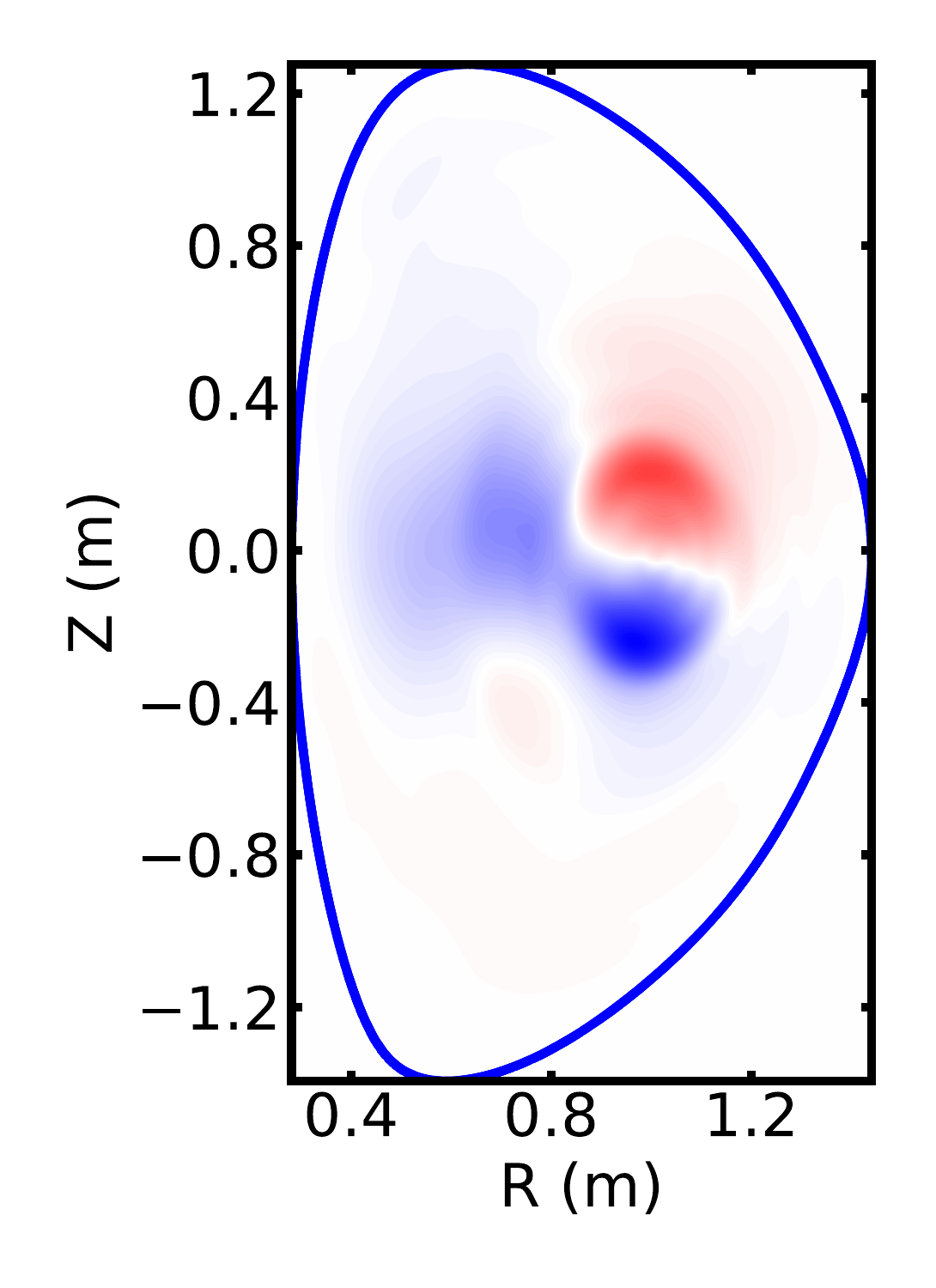}
			\put(50,20) {\textsf{(a)} $\delta \phi$}
		\end{overpic}
		\begin{overpic}[width=0.23\textwidth,trim=95 0 0 0,clip]{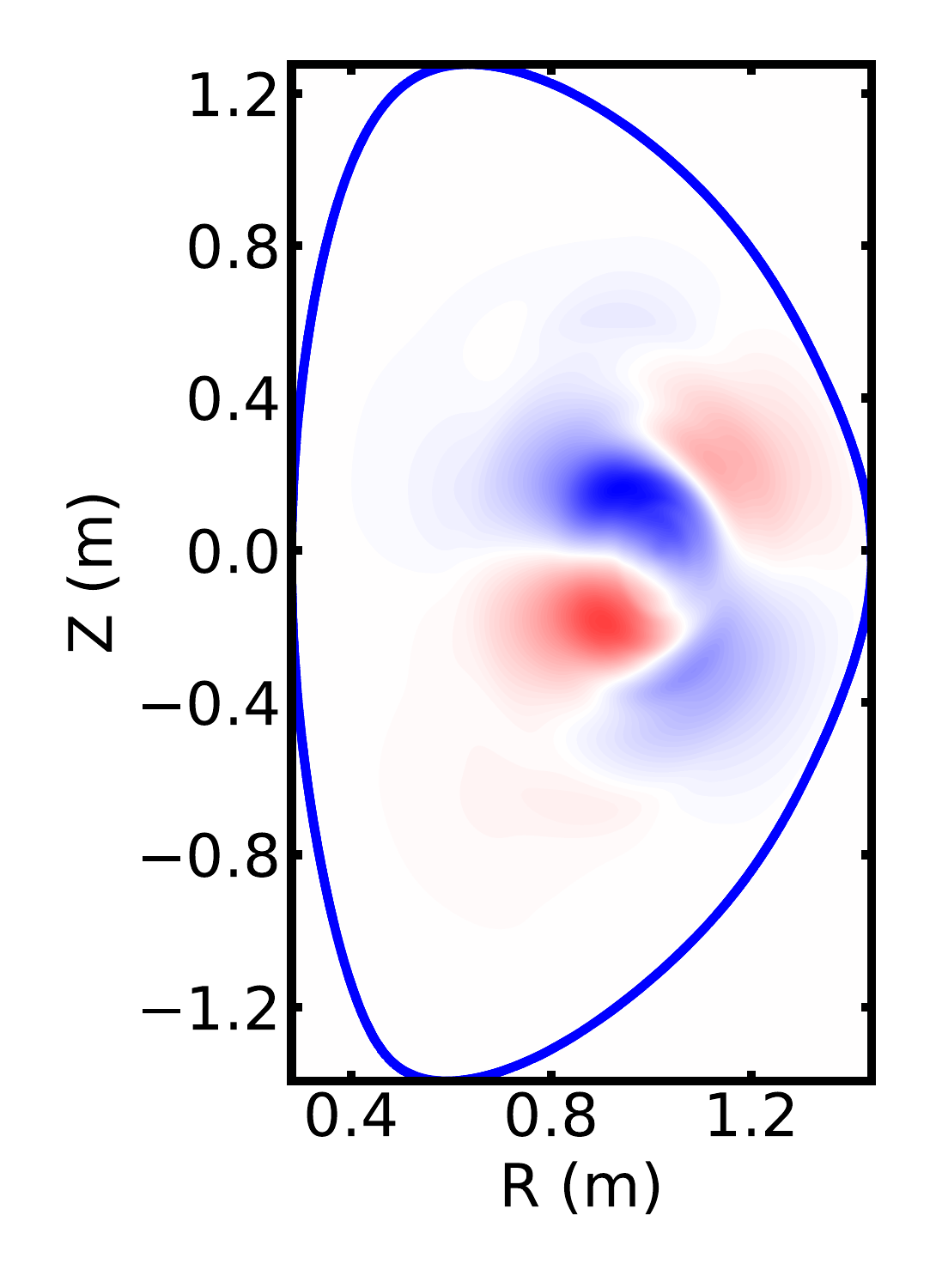}
			\put(28,20) {\textsf{(b)} $\delta \psi$}
		\end{overpic}
		\includegraphics[width=0.09\textwidth,trim=285 0 0 0,clip]{colorbar}
	\end{center}
	\caption{2D structure of $\delta\phi$ (a) and $\delta\psi$ (b) from NSTX $n=1$ linear simulation of the $q_\mathrm{min}=1.12$ case with thermal ions.}
	\label{fig:nstxphi}
\end{figure}

After including the thermal ion kinetic effects, the frequencies of fishbone-like modes increase significantly, while the growth rates decrease because of Landau damping. It is known that the fishbone mode is driven by the trapped ions which can have resonances with the mode through precession motion. However, as shown in \cref{fig:f-contour}, the population of trapped particles ($\lambda\approx 0$) in fast ions is relatively small compared to the co-passing ions, which leads to the low frequency of the fishbone-like modes. The thermal ions, on the other hand, have an isotropic distribution and can provide a large population of trapped ions to drive the fishbone-like mode. The mode structure of the fishbone-like mode for $q_\mathrm{min}=1.12$ is shown in \cref{fig:nstxphi}. For $q_\mathrm{min}>1.22$, the Landau damping effect stabilizes the \gls{bae}-like mode, which is similar to the DIII-D results.

In the NSTX experiment, both the kink ($m=1$) and tearing ($m=2$) modes have been identified using soft x-ray diagnostics, and the synergy effect of the two modes can lead to fast ion transport\citep{yang_synergy_2021}. It is believed that the $m=2$ mode is mainly neoclassical tearing mode (NTM) which is affected by the current associated with \glspl{ep}. The $m=1$ mode is marginally unstable in M3D-C1-K simulation with thermal ions under the experimental condition ($1.1<q_\mathrm{min}<1.2$), similar to the DIII-D case, which may be a result of pressure and current relaxation due to the nonlinear saturation of fishbone mode. The frequency observed in experiment is less than 5kHz after subtracting the Doppler frequency of toroidal rotation, which is close to the simulation frequency with $q_\mathrm{min}=1.12$. Note that the oscillation frequency after nonlinear saturation can be lower than the linear result due to the down-chirping of fishbone mode.

\section{Nonlinear simulation of fishbone-like mode in NSTX}
\label{sec:nstx-nonlinear}

Based on the linear simulation, we conduct the nonlinear simulation of the $n=1$ fishbone-like mode including the thermal ion kinetic effects in NSTX. We use the same \gls{mhd} equilibrium and \gls{ep} distribution as in \cref{sec:nstx-fishbone}, with $q_\mathrm{min}=1.12$. The simulation utilized a 3D finite element mesh, and nonlinear terms are included in the \gls{mhd} and $\delta f$ equations. The mesh has 8 toroidal planes connected by Hermite finite elements in the toroidal direction, which is good enough to resolve the $n=1$ perturbation, as we focus on the growth and saturation of the $n=1$ mode and its nonlinear coupling with the $n=0$ mode in this simulation. The mesh structure of each plane is the same as in the linear simulation. We use $16\times 10^6$ particle markers for the \gls{pic} simulation.

The nonlinear simulation was conducted at Perlmutter cluster at NERSC. For each nonlinear simulation we use 8 nodes with 64 AMD cores and 4 NVIDIA Tesla A100 GPUs on each node. The GPUs are used to do the MHD equation matrix element calculation and particle pushing. The CPU was used to solve the matrix by calling PETSC library, and calculate the particle moments.

\begin{figure}
	\begin{center}
		\begin{overpic}[width=0.45\textwidth]{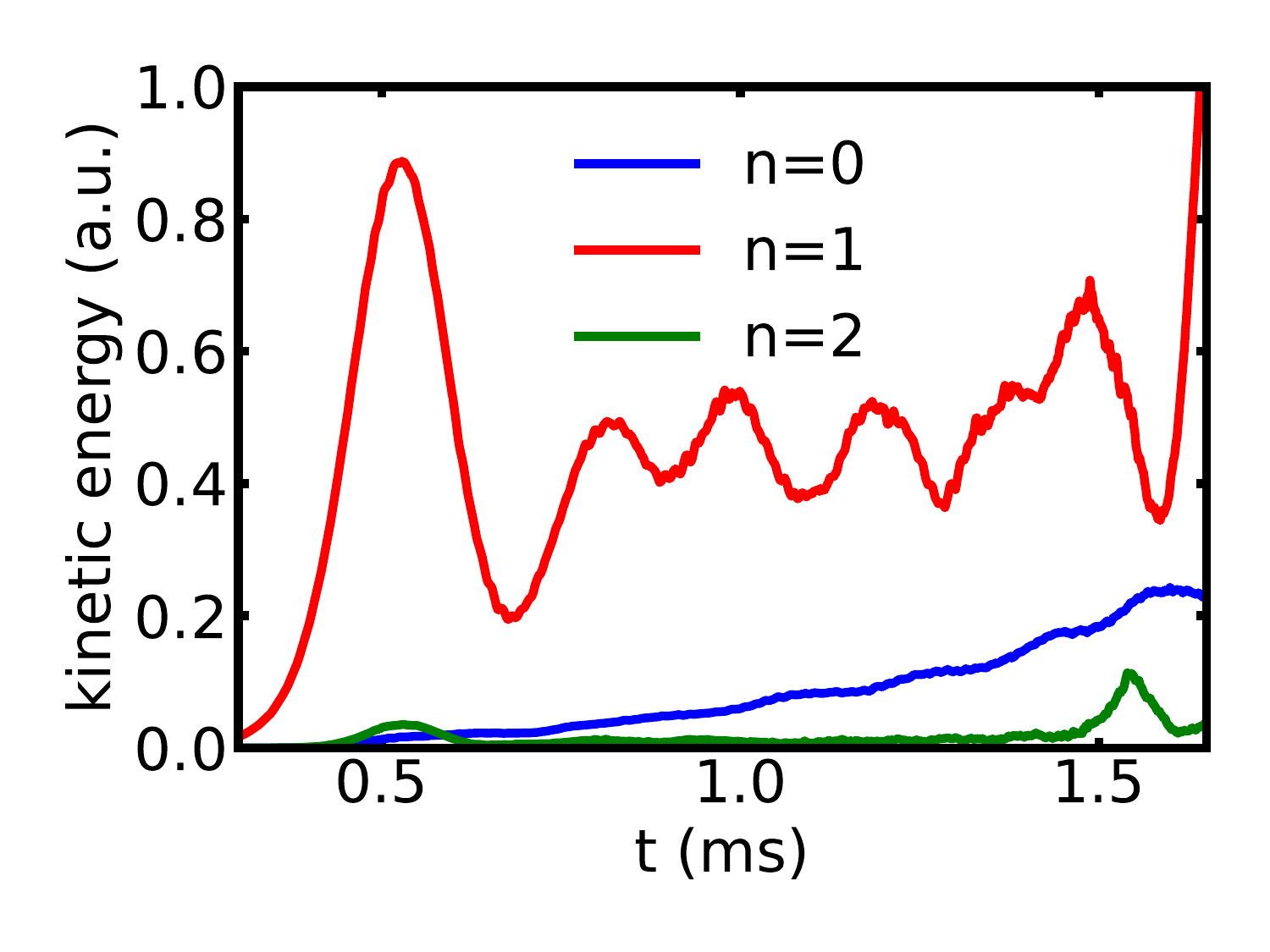}
			\put(83,20) {\textsf{(a)}}
		\end{overpic}
		\begin{overpic}[width=0.45\textwidth]{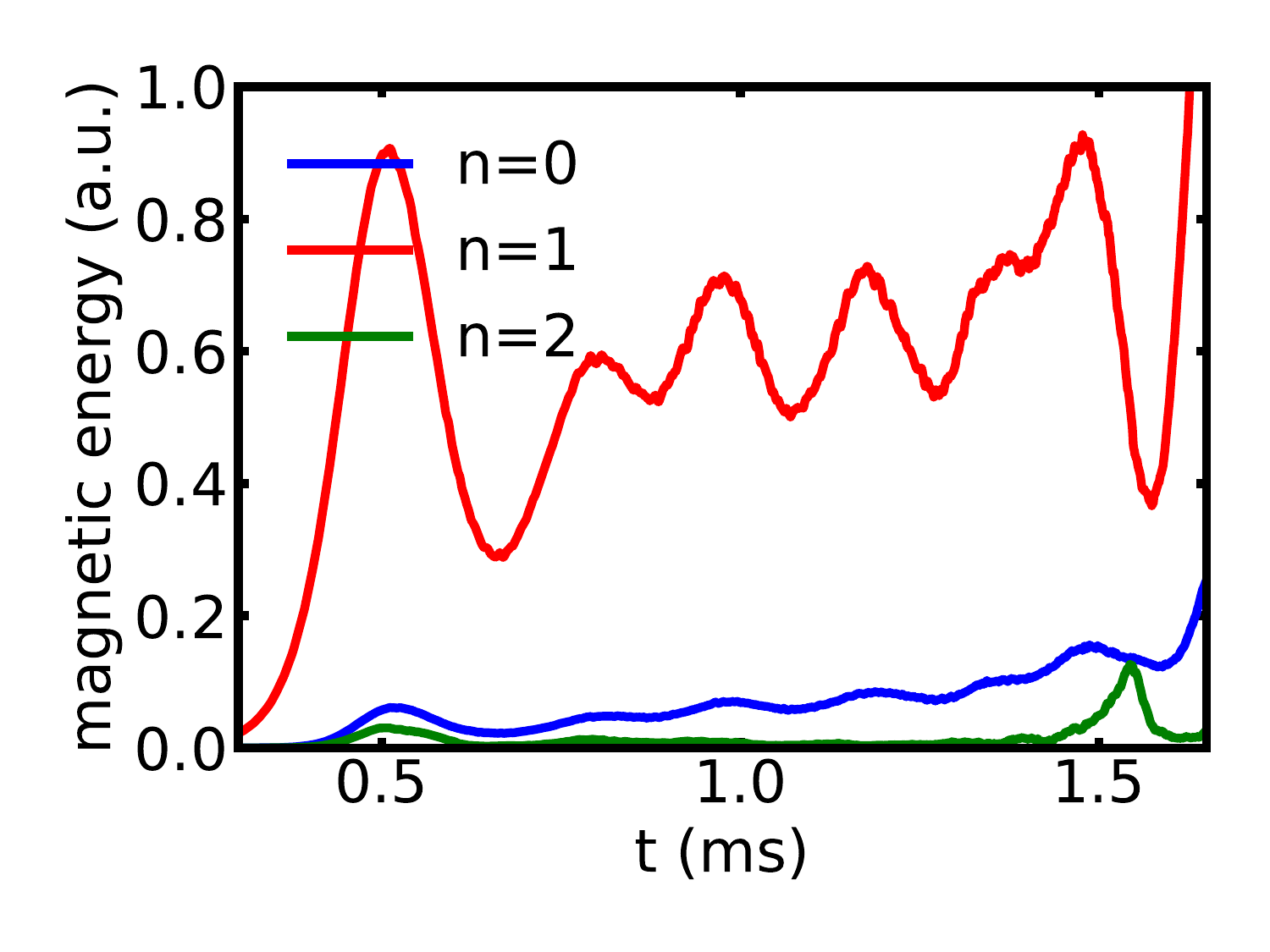}
			\put(83,22) {\textsf{(b)}}
		\end{overpic}
	\end{center}
	\caption{Time evolution of kinetic energy (a) and magnetic energy (b) of different toroidal harmonics from NSTX nonlinear simulation of the $q_\mathrm{min}=1.12$ case with thermal ions.}
	\label{fig:energy-evolution}
\end{figure}

\begin{figure}
	\begin{center}
		\begin{overpic}[width=0.33\textwidth]{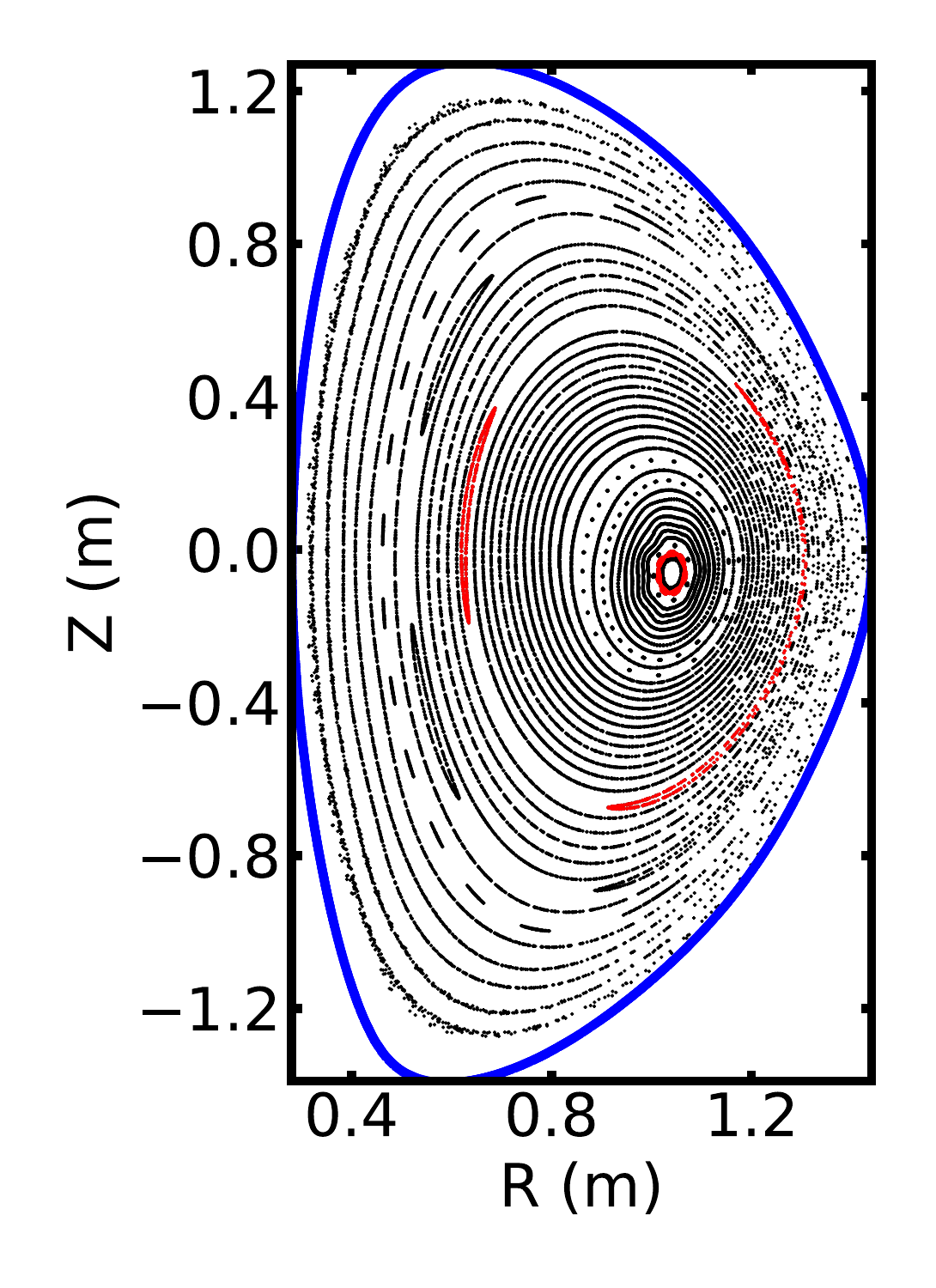}
			\put(58,18) {\textsf{(a)}}
		\end{overpic}
	    \raisebox{0.16\height}{\begin{overpic}[width=0.45\textwidth]{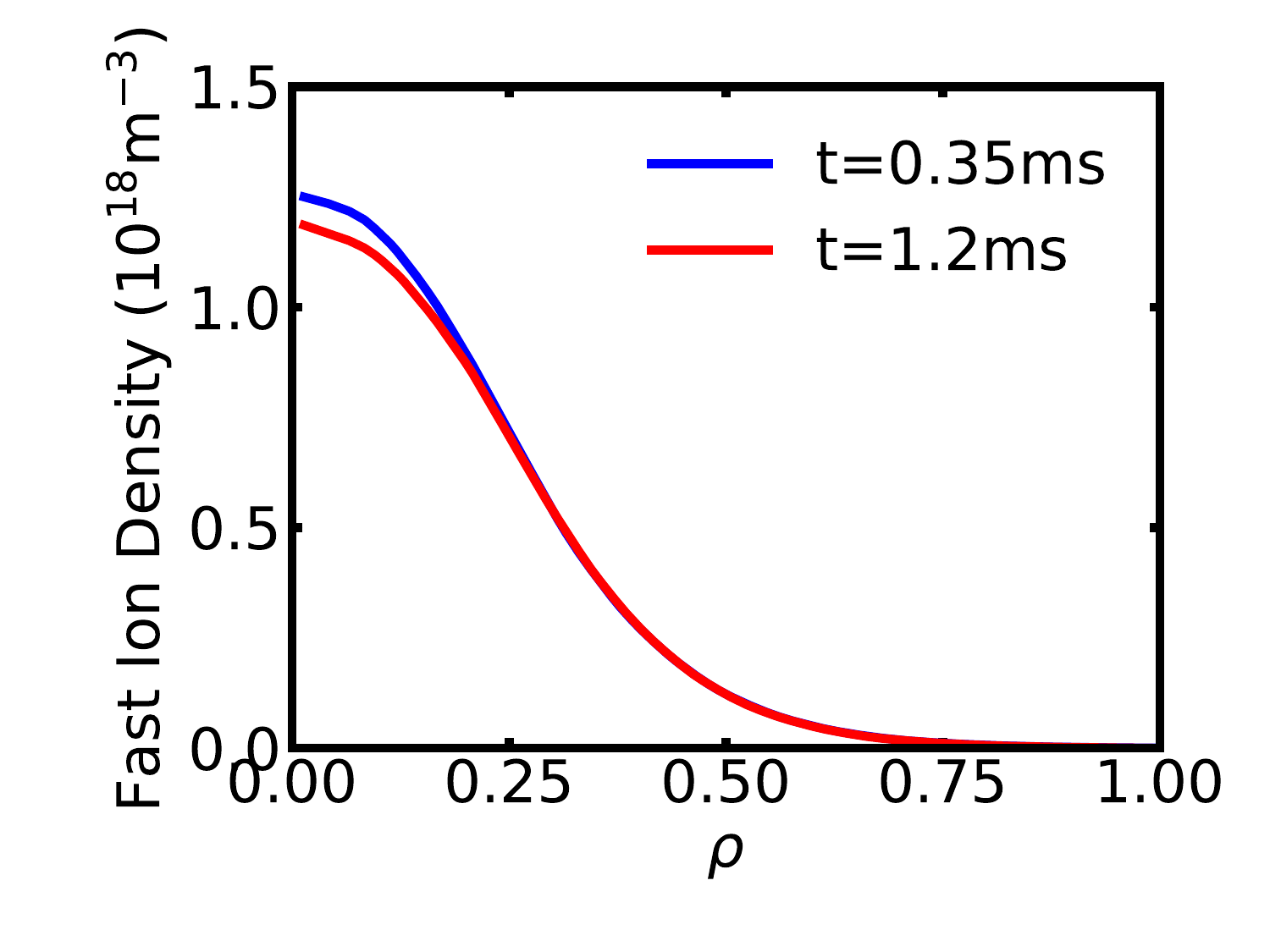}
	    \put(80,20) {\textsf{(b)}}
		\end{overpic}}
	\end{center}
	\caption{(a) Poincaré plot of magnetic flux surfaces at $t=1.2$ms. (1,1) and (2,1) islands are marked as red. (b) Change of fast ion density profile in the nonlinear simulation of $q_\mathrm{min}=1.12$ due to mode excitation.}
	\label{fig:poincare}
\end{figure}

The time evolution of the kinetic and magnetic parts of the MHD energy in the nonlinear NSTX fishbone simulation is shown in \cref{fig:energy-evolution}. We find that the $n=1$ mode reaches a saturation point at about $t=0.5$ms. The peak $\delta B/B_0$ at this point is about $1.5\times 10^{-2}$. There is also a $n=0$ mode excited due to the nonlinear mode-mode coupling. After the saturation, the magnetic energy of the $n=1$ mode stays at a high level, and the energy of $n=0$ mode keeps growing, meaning that the change to the magnetic field topology does not decay. The Poincaré plot of magnetic flux after mode saturation ($t=1.2$ms) is shown in \cref{fig:poincare} (a). There is a clear shift of the magnetic axis due to the excitation of (1,1) mode. The kink boundary overlaps with the flux contour of $q=q_\mathrm{min}$, meaning that most of the magnetic perturbation happens in the reversed-shear region. Although the equilibrium has $q>1.12$, the excitation of the mode creates a (1,1) island near the magnetic axis, and (2,1) islands near the $q=2$ surface, which is consistent with the experimental observation.

The growth of the $n=1$ mode can lead to transport of \glspl{ep}. \cref{fig:poincare} (b) shows the flux-averaged profiles of \gls{ep} density before and after mode saturation. This result shows that although the $n=1$ mode can provide a large $\delta B$ field, it only leads to a slight drop of \gls{ep} density (about 5\%) near the axis. The profile gradient does not drop to zero after saturation because the mode is susceptible to Landau damping, which means that is requires a finite radial gradient to excite. However, \cref{fig:poincare} (b) is a flux-averaged result and does not show the change of gradient along the mode's resonance line in phase space, which may be more significant.

\section{Summary}
\label{sec:summary}

The new kinetic-MHD simulation approach in this paper includes all the ions as kinetic particles, which is different from the classical approach that only deals with kinetic effects of fast ions. To implement that, part of the \gls{mhd} equations, the density equation and the parallel velocity equation, are replaced by synchronization from kinetic particle simulation, to avoid redundant calculation and parasitic modes caused by numerical errors. The rest of the \gls{mhd} equations can still be solved using a semi-implicit method with large timestep, with is different from the fully-kinetic or gyrokinetic simulation approach. The inclusion of thermal ion kinetic effects is important for macroscopic instability simulations targeting ITER and fusion reactors with large ion temperature.

The new simulation method has been used to study the thermal ion Landau damping of \glspl{iaw} and fishbone modes. It is found that the $n=1$ fishbone modes driven in an equilibrium with $q_\mathrm{min}>1$, or non-resonant fishbone modes, can be strongly affected by the Landau damping effects, since the mode phase velocity in the parallel direction is of the same order as the ion thermal velocity. In the linear simulation using a DIII-D and NSTX equilibrium, it is found that the $n=1$ \gls{bae}-like modes driven by fast ions can be stabilized by the thermal ions. Therefore it is necessary to revisit those non-resonant fishbone simulations done by kinetic-MHD codes, and add the thermal ion kinetic effects.

In developing the kinetic-MHD method, we include the parallel electric field calculated from electron pressure in the kinetic equations, which is essential for the \gls{iaw} simulation. This term however is not included in the \gls{mhd} equations. The two-fluid terms, including the parallel and perpendicular electric fields driven by Hall terms and electron pressure, can be important for the calculation of plasma waves with wavelength comparable to ion skin depth, such as \glspl{kaw} and whistler waves. The two-fluid terms have been developed in M3D-C1 and used to calculate their effects on magnetic reconnection\citep{beidler_local_2016}, but the simulation with them requires further testing and improvement of the matrix solver. We plan to do simulations with electric fields in both fluid and kinetic equations, to study the two fluid effects self-consistently in future work.

\section*{Acknowledgments}

We would like to thank Yasushi Todo, Masahiko Sato, Ruirui Ma, Feng Wang, Xiaolong Zhu, Zhihong Lin, Guillaume Brochard, Ge Dong and Chalson Kim for fruitful discussion. This work was supported by US Department of Energy grants DE-AC02-09CH11466. This research used the Traverse cluster at Princeton University. This research also used the Perlmutter cluster of the National Energy Research Scientific Computing Center (NERSC), using a NERSC NESAP Tier 2 award.

\bibliographystyle{jpp}

\bibliography{paper}

\end{document}